\documentclass[%
 reprint,
 superscriptaddress,
 preprintnumbers,
 nofootinbib,
 amsmath,amssymb,
 aps,
]{revtex4-2}
\usepackage{soul}
\usepackage{graphicx}
\usepackage{dcolumn}
\usepackage{bm}
\usepackage{hyperref}
\hypersetup{
	colorlinks=true,
	linkcolor=blue,
	filecolor=magenta,      
	urlcolor=blue,
	citecolor=magenta
}
\usepackage{amsmath,amssymb,mathtools}
\usepackage{graphics,graphicx,tikz,tensor}
\usetikzlibrary{arrows,decorations.pathmorphing,patterns}

\begin{document}

\preprint{QMUL-PH-24-03}

\title{Post-Newtonian Multipoles from the Next-to-Leading \\ Post-Minkowskian Gravitational Waveform}

\author{Alessandro Georgoudis}\email{a.georgoudis@qmul.ac.uk}
\affiliation{Centre for Theoretical Physics, Department of Physics and Astronomy, \\ Queen Mary University of London, Mile End Road, London E1 4NS, United Kingdom}
\author{Carlo Heissenberg}\email{c.heissenberg@qmul.ac.uk}
\affiliation{School of Mathematical Sciences, Queen Mary University of London, \\ Mile End Road, London, E1 4NS, United Kingdom}
\author{Rodolfo Russo}\email{r.russo@qmul.ac.uk}
\affiliation{School of Mathematical Sciences, Queen Mary University of London, \\ Mile End Road, London, E1 4NS, United Kingdom}



\begin{abstract}
We consider the frequency-domain LO and NLO post-Minkowskian (PM) waveforms obtained from the tree-level and one-loop amplitudes describing the scattering of two massive scalar objects and the emission of one graviton. We explicitly calculate their post-Newtonian (PN) limit
obtaining an expansion up to the third subleading PN order in all ingredients: the tree-level amplitude, the odd and even parts of the real one-loop kernel,
and the Compton or ``rescattering'' cuts, 
thus reaching 3PN precision for the latter.
We provide explicit expressions for the multipole decomposition of these results in the center-of-mass frame and compare them with the results obtained from the classical Multipolar post-Minkowskian (MPM) method. We find perfect agreement between the two, once the BMS supertranslation frame is properly adjusted and the infrared divergences due to rescattering are suitably subtracted in dimensional regularization. This shows that the approach proposed in~\cite{Georgoudis:2023eke} can be applied beyond the soft-regime ensuring the agreement between amplitude-based and MPM results for generic frequencies.
\end{abstract}

\maketitle


\section{Introduction.} 

Amplitude-based methods have been instrumental in recent years in improving the state of the art of post-Minkowskian (PM) calculations, leading in particular to new results for the deflection angle characterizing hyperbolic gravitational encounters at large impact parameters \cite{Bern:2019nnu,Bern:2019crd,Bern:2021yeh,Bern:2021dqo}. These achievements were based on multiloop calculations of the elastic $2\to2$ amplitude for the scattering of two minimally-coupled massive scalars, which model colliding Schwarzschild black holes. While initially the focus was on the so-called conservative sector of the interactions \cite{Cheung:2018wkq,KoemansCollado:2019ggb,Cheung:2020gyp,Cristofoli:2020uzm}, it was soon realized that the physical result for the deflection must also include dissipative effects, such as radiation-reaction and recoil, which can be captured by taking into account suitable cuts involving also on-shell gravitons \cite{DiVecchia:2020ymx,Damour:2020tta,DiVecchia:2021ndb,DiVecchia:2021bdo,Herrmann:2021tct,Bjerrum-Bohr:2021vuf,Bjerrum-Bohr:2021din,Damgaard:2021ipf,Brandhuber:2021eyq,Damgaard:2023vnx,Damgaard:2023ttc}. Outstanding achievements were also obtained by means of quantum-field-theory inspired worldline techniques \cite{Kalin:2020mvi,Kalin:2020fhe,Mogull:2020sak,Dlapa:2021npj,Jakobsen:2022psy,Kalin:2022hph,Dlapa:2022lmu,Dlapa:2023hsl} and closely related methods were developed to calculate various kinds of integrated radiative observables as well \cite{Herrmann:2021lqe,Manohar:2022dea,DiVecchia:2022owy,Mougiakakos:2022sic,Riva:2022fru,DiVecchia:2022nna,Heissenberg:2022tsn,Heissenberg:2023uvo}. Many results have been analytically continued from hyperbolic to elliptic trajectories~\cite{Kalin:2019rwq,Kalin:2019inp,Saketh:2021sri,Cho:2021arx}, showing that it is possible to use amplitude-based methods to extract information about bound systems as well.

The success of this program, which is mainly motivated by the dawn of a high-precision era of gravitational-wave observations \cite{Purrer:2019jcp,Maggiore:2019uih,Barausse:2020rsu,Ballmer:2022uxx}, sparked renewed interest in another observable associated to hyperbolic encounters: the  gravitational scattering waveform.
The leading PM result for this quantity dates back to the seminal works of Kovacs and Thorne \cite{Kovacs:1977uw,Kovacs:1978eu}, which were recently revisited and streamlined using worldline methods in \cite{Jakobsen:2021smu,Mougiakakos:2021ckm}  and amplitude methods in \cite{DiVecchia:2021bdo,DeAngelis:2023lvf,Brandhuber:2023hhl,Aoude:2023dui}.
In the latter approach, the leading PM waveform arises as the Fourier transform of the tree-level $2\to3$ amplitude with one graviton emission  \cite{Cristofoli:2021vyo}, whose explicit expression is given in \cite{Goldberger:2016iau,Luna:2017dtq}.
The dynamical information needed to calculate the first subleading correction to this result is instead encoded in the one-loop $2\to3$ amplitude that was calculated in \cite{Brandhuber:2023hhy,Herderschee:2023fxh,Elkhidir:2023dco,Georgoudis:2023lgf}. This subleading PM object contains several interesting new features. 

First, it involves ``superclassical'' or iteration terms, which are expected from the eikonal exponentiation \cite{Cristofoli:2021jas,Damgaard:2021ipf,DiVecchia:2022piu} but cannot appear in the final result for a physical observable, and were thus simply dropped in the original series of works.
It was later pointed out in \cite{Caron-Huot:2023vxl}, upon closer inspection of the Kosower--Maybee--O'Connell (KMOC) prescription for calculating observables from amplitudes \cite{Kosower:2018adc}, that the cancellation of such superclassical terms actually leaves behind a classical contribution arising from the difference between two-massive-particle cuts, which we shall denote by $\frac{i}{2}(s-s')$. However, we showed in \cite{Georgoudis:2023eke} that the omission of this term is simply compensated by rotating the particles' velocities and impact parameter from the initial to the ``eikonal'' ones (up to a shift in the retarded time). This is equivalent to switching from the reference frame anchored to the initial velocities to the ``eikonal'' frame, which are rotated by $\Theta/2$ with respect to each other, $\Theta$ denoting the PM deflection angle. Working directly with the latter, one can thus omit $\frac{i}{2}(s-s')$ and focus on the classical part of the amplitude in \cite{Brandhuber:2023hhy,Herderschee:2023fxh,Georgoudis:2023lgf}. This mechanism had already been observed in \cite{Bini:2023fiz} in the post-Newtonian (PN) limit and in \cite{Aoude:2023dui} in the soft limit. The contribution $\frac{i}{2}(s-s')$ was also calculated in momentum space and included in the revised versions of \cite{Brandhuber:2023hhy,Herderschee:2023fxh} and in \cite{Georgoudis:2023ozp,Bohnenblust:2023qmy}.

Second, the subleading waveform kernel involves an infrared divergent part that arises due to tail or rescattering effects
\cite{Blanchet:1987wq,Blanchet:1992br,Blanchet:1993ng}
, which can be however exponentiated into a (divergent) phase factor \cite{Weinberg:1965nx}. While this can be reabsorbed into a shift or ``renormalization'' of the observer's retarded time \cite{Goldberger:2009qd,Porto:2012as}, an open issue remained as to whether this exponentiation also leaves behind a correction in the finite piece. We shall see that this is indeed the case as such contributions arise from $\epsilon/\epsilon$ corrections that must be taken into account when taking the $\epsilon\to0$ limit in dimensional regularization, $D=4-2\epsilon$.

The main motivation for the present work is to clarify the comparison between the amplitude-based result for the waveform up to subleading PM order and the predictions obtained from more conventional methods for waveform calculations employed in classical General Relativity (GR). These are based on the PN solution of the Einstein equations in the near zone, followed by a matching with the multipole expansion of the gravitational field in the exterior zone and its propagation to the far zone, where it determines the radiative multipoles
\cite{Blanchet:1985sp,Blanchet:1986dk,Blanchet:1998in,Poujade:2001ie}
(see \cite{Blanchet:2013haa} for a comprehensive review). The comparison in the PN limit between  the amplitude waveform and the Multipolar post-Minkowskian (MPM) one  was initiated in \cite{Bini:2023fiz}. There, the authors found agreement up to 1PN for the real part of the one-loop kernel $\mathcal B$ and 1.5PN for the imaginary part $\frac{i}{2}(c_1+c_2)$, but found a mismatch at 2.5PN order in the latter. When re-expanded for small frequencies $\omega\to0$, this discrepancy would start to appear at order $\omega\log\omega$.

We later reassessed this disagreement in \cite{Georgoudis:2023eke}, focusing on the soft limit $\omega\to0$. There, after checking the agreement between the amplitude result and predictions coming from soft theorems \cite{Sahoo:2018lxl,Sahoo:2021ctw} for the universal terms, we also calculated the first non-universal order, $\omega\log\omega$, at one loop and showed that all ``mismatches'' up to and including 2.5PN could be in fact reabsorbed by adjusting the origin of retarded time and the choice of Bondi--Metzner--Sachs (BMS) frame. This highlighted once again \cite{Veneziano:2022zwh} the important role played by BMS supertranslations when comparing amplitude-based results, which, for $\omega>0$, are more naturally expressed in the ``canonical'' frame where the asymptotic shear vanishes in the far past, and MPM formulas, which only hold in the ``intrinsic'' frame as they crucially rely on the $\mathcal O(G)$ time-independent part of the shear due to free motion.

In this work, we extend and complement the analysis of \cite{Bini:2023fiz,Georgoudis:2023eke} by calculating the amplitude waveform in the PN limit without resorting to the soft approximation.
We perform the expansion for small velocity in momentum-space, the Fourier transforms and the multipolar decomposition up to the third subleading order (NNNLO) in the PN expansion for all the basic ingredients of the waveform kernel. These are the tree-level amplitude, the odd and even part of the real kernel $\mathcal B$, and the Compton cuts $\frac{i}{2}(c_1+c_2)$. We collect our results in the ancillary file in a computer-friendly format. We then briefly recall the MPM techniques and use them to provide independent predictions up to the desired order in the velocity. Finally, we compare the two sets of results, finding perfect agreement once two aspects are appropriately taken into account.

The first one is the BMS supertranslation already highlighted in \cite{Veneziano:2022zwh,Georgoudis:2023eke}, which maps from the amplitude (canonical) to the MPM (intrinsic) supertranslation frame. The second one is the subtraction of finite $\epsilon/\epsilon$ terms that is determined by the exponentiation of infrared divergences. In particular, the NNLO PN calculation of the Compton cuts beyond the soft limit is the simplest nontrivial comparison that is able to nail down this subtraction, because up to soft order $\omega\log\omega$ such terms could be also reabsorbed by adjusting the origin of retarded time, while this is no longer true for generic $\omega$. 
This mechanism was already noted in \cite{Georgoudis:2023eke} when discussing the $\frac{i}{2}(s-s')$ contribution for generic frequency and velocity. It was instead immaterial for the analysis of the Compton cuts performed there, which was restricted to the non-analytic terms in the soft limit.

The paper is organized as follows. In Section~\ref{sec:reminder}, we summarize the current understanding of the subleading waveform kernel derived from amplitudes and describe its main building blocks. In Section~\ref{sec:IRdiv}, we discuss the exponentiation of infrared divergences, taking into account the resulting subtraction of finite $\epsilon/\epsilon$ terms.  Section~\ref{sec:PNampl} then deals with the PN expansion of the amplitude-based waveform, for which we describe the main steps. For simplicity, we provide the explicit expressions only of various contributions to the quadrupole in the text, while collecting the entire set of multipoles in the ancillary file.
In Section~\ref{sec:MPM}, we discuss instead the MPM method, reviewing how time-reversal even and odd effects arise in that framework, and illustrating in particular tail, nonlinear and radiation-reaction effects.
We conclude in Section~\ref{sec:comparison} by comparing the two methods, explaining how agreement is achieved and highlighting along the way the strengths and drawbacks of either approach.
In Appendix~\ref{sec:multipoles}, we provide for completeness a self-contained illustration of the multipole expansion and collect some useful properties of symmetric trace-free (STF) tensors.

\paragraph*{Note added.}
At the final stages of this work, we became aware of parallel work by Donato Bini, Thibault Damour, Stefano De Angelis, Andrea Geralico, Aidan Herderschee, Radu Roiban and Fei Teng, which partly overlaps with ours. We thank them for communication and for coordinating the submissions.

\section{Reminder on the Waveform Kernel up to One Loop}
\label{sec:reminder}

We begin by recalling that the metric fluctuation $h_{\mu\nu}(x)$, defined by 
\begin{equation}\label{}
	g_{\mu\nu}(x)-\eta_{\mu\nu} = h_{\mu\nu}(x),
\end{equation}
can be expressed as follows in terms of the \emph{wave-shape} $\tilde{W}_{\mu\nu}$,
\begin{equation}\label{}
	h_{\mu\nu}(x) = 2\kappa\int_k \left[
	e^{ik\cdot x} \tilde{W}_{\mu\nu}(k)
	+
	e^{-ik\cdot x} \tilde{W}_{\mu\nu}^\ast(k) \right],
\end{equation}
with $\kappa = \sqrt{8\pi G}$ and $\int_k=\int \frac{d^4k}{(2\pi)^4}\,2\pi\theta(k^0)\delta(k^2)$ the on-shell phase-space graviton measure. In the asymptotic limit $r\to\infty$ for a fixed retarded time $U$ and a given null vector $n^\mu$, one obtains the following \emph{formal} expression for the $D=4$ asymptotic waveform \cite{Cristofoli:2021vyo,DiVecchia:2023frv},
\begin{equation}\label{eq:naivh}
	h_{\mu\nu}(x) \sim \frac{4G}{\kappa r}\int_0^\infty
	 e^{-i\omega U} \tilde{W}_{\mu\nu}(\omega n)
	 \,
	 \frac{d\omega}{2\pi}
	+
 (\text{c.c.})
\end{equation}
up to  
corrections that are further suppressed for large $r$ (``c.c.'' stands for ``complex conjugate'').
The expression \eqref{eq:naivh} is formal indeed, since at one loop one cannot take the $D=4$ limit in a straightforward way owing to the presence of infrared divergences. We shall see below how their exponentiation allows one to arrive at a precise and concrete prescription for the $D=4$ observable.

As discussed in  Refs.~\cite{Brandhuber:2023hhy,Herderschee:2023fxh,Elkhidir:2023dco,Georgoudis:2023lgf,Caron-Huot:2023vxl}, in the KMOC approach \cite{Kosower:2018adc}, the wave-shape $\tilde{W}_{\mu\nu}$ can be expressed as the Fourier transform of a momentum-space ``waveform kernel'' $W_{\mu\nu}$, 
\begin{equation}\label{eq:FT5}
	\begin{split}
		&\tilde W_{\mu\nu}(k)
		\! = \! \! \int \frac{d^D q_1}{(2\pi)^D} \frac{d^D q_2}{(2\pi)^D}\,2\pi\delta(2p_1\cdot q_1)\,2\pi\delta(2p_2\cdot q_2)\\
		&\times e^{iq_1\cdot b_1+i q_2\cdot b_2}\,W_{\mu\nu}(q_1,q_2)\,(2\pi)^D\delta^{(D)}(q_1+q_2+k)\,.
	\end{split}
\end{equation}
Here we adopted the following notation for the momenta of the incoming and outgoing states entering the amplitude, summarized in the picture below,
\begin{equation}\label{}
	\begin{gathered}
		\begin{tikzpicture}[scale=.7]
			\path [draw, ultra thick, blue] (-4.5,2.2)--(-1.5,2.2);
			\path [draw, ultra thick, green!60!black] (-4.5,.8)--(-1.5,.8);
			\path [draw, red] (-3,1.5)--(-1.5,1.5);
			\filldraw[white] (-3,1.5) ellipse (.7 and 1);
			\filldraw[pattern=north east lines, thick] (-3,1.5) ellipse (.7 and 1);
			\draw[thick] (-3,1.5) ellipse (.7 and 1);
			\node at (-4.5,2.2)[left]{$p_1$};
			\node at (-1.5,2.2)[right]{$p_4=q_1-p_1$};
			\node at (-4.5,.8)[left]{$p_2$};
			\node at (-1.5,.8)[right]{$p_3=q_2-p_2$};
			\node at (-1.5,1.5)[right]{$k$};
		\end{tikzpicture}
	\end{gathered}
\end{equation}
where all external momenta are treated as formally outgoing.
Up to subleading order in the PM expansion, the KMOC kernel is given by $W^{\mu\nu}= W_{0}^{\mu\nu}+W_1^{\mu\nu}+\cdots$ with
\begin{align}\label{eq:W0W1explN}
	\begin{split}
	W_0^{\mu\nu} &= \mathcal A_0^{\mu\nu} \,,
	\\ 
	W_1^{\mu\nu} &= \mathcal B_1^{\mu\nu} +\frac{i}{2} (s^{\mu\nu}-s'^{\mu\nu}) + \frac{i}{2}(c_1^{\mu\nu}+c_2^{\mu\nu})\,.
	\end{split}
\end{align}
Here, $\mathcal{A}_0$ is the tree-level $2\to3$ amplitude in the classical limit and $\mathcal B_{1}$ is the real part of the one-loop $2\to3$ amplitude (the $N$-matrix element). Instead, $s$, $s'$, $c_1$ and $c_2$ denote the one-loop unitarity cuts, which can be depicted schematically as follows,
\begin{equation}\label{eq:Schannel}
	s=
	\begin{gathered}
		\begin{tikzpicture}[scale=0.75]
			\path [draw, ultra thick, blue] (-4,2)--(-.3,2);
			\path [draw, ultra thick, green!60!black] (-4,1)--(-.3,1);
			\path [draw, red] (-1,1.5)--(-.32,1.5);
			\filldraw[black!20!white, thick] (-3,1.5) ellipse (.5 and .8);
			\draw[thick] (-3,1.5) ellipse (.5 and .8);
			\filldraw[black!20!white, thick] (-1.3,1.5) ellipse (.5 and .8);
			\draw[thick] (-1.3,1.5) ellipse (.5 and .8);
		\end{tikzpicture}
	\end{gathered}
 \quad
	s'=
	\begin{gathered}
		\begin{tikzpicture}[scale=0.75]
			\path [draw, ultra thick, blue] (-4,2)--(-.3,2);
			\path [draw, ultra thick, green!60!black] (-4,1)--(-.3,1);
			\path [draw, red] (-3,1.5)--(-2.1,1.5);
			\filldraw[black!20!white, thick] (-3,1.5) ellipse (.5 and .8);
			\draw[thick] (-3,1.5) ellipse (.5 and .8);
			\filldraw[black!20!white, thick] (-1.3,1.5) ellipse (.5 and .8);
			\draw[thick] (-1.3,1.5) ellipse (.5 and .8);
		\end{tikzpicture}
	\end{gathered}
\end{equation}
and
\begin{equation}\label{eq:Cchannel}
	c_1=
	\begin{gathered}
		\begin{tikzpicture}[scale=0.75]
			\path [draw, ultra thick, blue] (-4,2)--(-.3,2);
			\path [draw, ultra thick, green!60!black] (-4,1)--(-2.1,1);
			\path [draw, red] (-3,1.5)--(-.3,1.5);
			\filldraw[black!20!white, thick] (-3,1.5) ellipse (.5 and .8);
			\draw[thick] (-3,1.5) ellipse (.5 and .8);
			\filldraw[black!20!white, thick] (-1.3,1.75) ellipse (.45 and .55);
			\draw[thick] (-1.3,1.75) ellipse (.45 and .55);
		\end{tikzpicture}
	\end{gathered}
	\quad
	c_2=
	\begin{gathered}
		\begin{tikzpicture}[scale=0.75]
			\path [draw, ultra thick, blue] (-4,2)--(-2.1,2);
			\path [draw, ultra thick, green!60!black] (-4,1)--(-.3,1);
			\path [draw, red] (-3,1.5)--(-.3,1.5);
			\filldraw[black!20!white, thick] (-3,1.5) ellipse (.5 and .8);
			\draw[thick] (-3,1.5) ellipse (.5 and .8);
			\filldraw[black!20!white, thick] (-1.3,1.25) ellipse (.45 and .55);
			\draw[thick] (-1.3,1.25) ellipse (.45 and .55);
		\end{tikzpicture}
	\end{gathered}
\end{equation}

In \cite{Georgoudis:2023eke}, we showed that the two-massive-particle-cut contributions  \cite{Caron-Huot:2023vxl} appearing on the right-hand side of \eqref{eq:W0W1explN} can be interpreted as a simple rotation of the particles' velocities and impact parameter, plus a (divergent but immaterial) time shift associated to the Shapiro time delay.
Instead of adopting as basis vectors the initial velocities, 
\begin{equation}\label{}
	p_1^\mu = -m_1 v_1^\mu\,,\qquad p_2^\mu = -m_2 v_2^\mu
\end{equation}
and the initial impact parameter
\begin{equation}\label{}
	b_J^\mu = b^\mu_1-b_2^\mu\,,\qquad b_J\cdot v_1=b_J\cdot v_2 = 0\,,
\end{equation}
it is therefore convenient to describe the two-body system in terms of rotated, ``eikonal'' classical variables 
\begin{equation}\label{}
	\begin{split}
	\tilde p_1^\mu = \tilde m_1 \tilde u_1^\mu  = -p_1^\mu+\frac12 Q^\mu\,,\\
	\tilde p_2^\mu = \tilde m_2 \tilde u_2^\mu  = -p_2^\mu-\frac12 Q^\mu\,,
	\end{split}
\end{equation}
with $Q^\mu$ the (PM expanded) impulse and denoting by $Q$, $b$ the magnitudes of $Q^\mu$, $b_{e}^\mu$ (or $b_{J}^\mu$),\footnote{Since $b^2_e$ and $b^2_J$ would differ by $\mathcal O(G^2)$ terms, up to the order of interest here we can set $b^2_J \simeq b_e^2 = b^2$.}
\begin{equation}\label{eq:be}
	b_e^\mu = b_J^\mu - \left(
	\frac{\check{v}_1^\mu}{2 m_1}
	-
	\frac{\check{v}_2^\mu}{2 m_2}
	\right)
	b\,Q
\end{equation}
after Fourier transform.
In \eqref{eq:be}, we also introduced
\begin{equation}\label{}
	\check v_1^\mu = \frac{\sigma v_2^\mu-v_1^\mu}{\sigma^2-1}\,,\qquad
	\check v_2^\mu = \frac{\sigma v_1^\mu-v_2^\mu}{\sigma^2-1}\,,
\end{equation}
where $\sigma=-v_1\cdot v_2$ as in \eqref{eq:invariants} below.
Note that, by definition, the rotated variables also obey
\begin{equation}\label{}
	b_e\cdot \tilde u_1  = b_e\cdot \tilde u_2=0\,.
\end{equation}
This choice amounts to simply dropping the $\frac{i}{2}(s-s')$ contribution to the kernel,
working with
\begin{equation}\label{eq:Weik}
	W^\text{eik}(k) = \mathcal A_0 + \mathcal B_{1} + \frac{i}{2}\left(c_1+c_2\right) .
\end{equation}
Here and in the following, we adopt a shorthand notation for gauge-invariant amplitudes contracted with physical polarization tensors $\mathcal A =\varepsilon_{\mu\nu}\mathcal A^{\mu\nu} =  \varepsilon_\mu \mathcal A^{\mu\nu}\varepsilon_\nu$ with $\varepsilon^\mu$ the usual $D=4$ polarization vectors such that $\varepsilon\cdot k=0$, $\varepsilon\cdot \varepsilon=0$, $\varepsilon\cdot \varepsilon^\ast = 1$.

Let us turn to the description of the ingredients entering \eqref{eq:Weik}. 
For the tree-level amplitude $\mathcal A_0$ in the classical limit, first derived in
\cite{Goldberger:2016iau,Luna:2017dtq}, we employ in particular the explicit expression given in \cite{DiVecchia:2021bdo,DiVecchia:2023frv}. 
We shall denote the kinematic invariants as follows,
\begin{equation}\label{eq:invariants}
	\sigma = -v_1\cdot v_2\,,\quad 
	\omega_1 = -v_1\cdot k\,,\quad
	\omega_2 = -v_2\cdot k\,.
\end{equation}
In the real part of the one-loop kernel, $\mathcal B_1$, one can isolate an \emph{odd} and an \emph{even} part, $\mathcal B_1=\mathcal B_{1O}+\mathcal B_{1E}$, under $\omega_{1,2}\mapsto -\omega_{1,2}$. The former being fixed in terms of the tree-level amplitude by the simple formula 
$\mathcal B_{1O}=\mathcal B_{1O}^{(i)}+\mathcal B_{1O}^{(h)}$ with
\cite{Brandhuber:2023hhy,Herderschee:2023fxh,Elkhidir:2023dco,Georgoudis:2023lgf}
\begin{subequations}\label{eq:B1odd}
\begin{align}
	\label{eq:B1odd(i)}
\mathcal B_{1O}^{(i)}&= -\frac{\sigma\left(\sigma^2-\frac{3}{2}\right)}{(\sigma^2-1)^{3/2}}
\,\pi G E \omega\,\mathcal{A}_0\,,
\\
		\label{eq:B1odd(h)}
\mathcal B_{1O}^{(h)}&= \pi G E \omega\,\mathcal{A}_0\,,
\end{align}
\end{subequations} 
where $E$ and $\omega$ are the total incoming energy and the outgoing graviton's frequency in the center-of-mass frame.
Instead, the structure of the even part is more intricate and takes the following form
(see e.g.~the ancillary files of \cite{Herderschee:2023fxh})
\begin{equation}\label{eq:B1even}
	\mathcal B_{1E} =
	\left[ \frac{A_{\omega_1}^R}{\omega_1^2(q_2^2+\omega_1^2)^{7/2}}+\frac{A_{q_1}^R}{\omega_2^2\sqrt{q_1^2}} 
	\right] \frac{m_1^3m_2^2}{q_2^2\mathcal Q_1^4}
	+ 
	(1\leftrightarrow2)\,.
\end{equation}
In Eq.~\eqref{eq:B1even}, we are displaying explicitly only the structure involving $m_1^3 m_2^2$, while the other one is obtained by interchanging the labels 1 and 2 everywhere. The functions $A_{X}^R$, with $X \in \{\omega_1, q_1\}$, denote polynomial functions of the kinematic invariants $\sigma$, $\omega_1$, $\omega_2$, $q_1^2$, $q_2^2$, quadratic in $\varepsilon\cdot u_1$, $\varepsilon\cdot u_2$ and $\varepsilon \cdot q_2$ (we recall that $\varepsilon\cdot k=0$ and hence $\varepsilon\cdot q_1=-\varepsilon\cdot q_2$).  
Moreover, 
\begin{equation}\label{}
	\mathcal Q_1 =  (q_1^2-q_2^2)^2-4q_1^2\omega_1^2
\end{equation}
represents an apparent or ``spurious'' pole in the expression, which does not correspond to any actual singularity as one can explicitly check by taking the limit $\mathcal Q_1\to0$ and using the vanishing of the Gram determinant,
\begin{equation}\label{eq:gram}
	\mathrm{det}\left[\mathrm{Gram}(v_1,v_2,q_1,q_2,\varepsilon)\right]=0\,.
\end{equation}
This encodes the fact that $\varepsilon^\mu$ can be always expressed in $D=4$ as a linear combination of $v_1$, $v_2$, $q_1$, $q_2$.
Analogously, in the $1\leftrightarrow2$ contribution we will have
\begin{equation}\label{}
	\mathcal Q_2 =  (q_1^2-q_2^2)^2-4q_2^2\omega_2^2\,.
\end{equation}
Finally, say, the $c_1$ cut is given as follows, again letting $A_{X}^I$, with $X \in \{\omega_1, q_1, \omega_1\omega_2, q_1q_2\}$, denote polynomials in the invariants quadratic in the polarization vectors, (see e.g.~the ancillary files of \cite{Herderschee:2023fxh} for the explicit expressions)
\begin{widetext}
\begin{align}\label{eq:c1}
		\frac{i}{2}\,c_1 
		&= i G m_1 \omega_1\left(-\frac{1}{\epsilon} + \log\frac{\omega_1^2}{\mu^2_\text{IR}}\right)\left[\mathcal A_0\right]_{D=4}
		+
		i  m_1^3 m_2^2 \mathcal M^{m_1^3m_2^2}\,,\\
		\label{eq:calMm1m2}
		\begin{split}
		\mathcal M^{m_1^3m_2^2}
		&=
		\frac{A_\text{rat}^I}{q_1^2q_2^2(\sigma^2-1)\omega_1\omega_2^2(q_2^2+\omega_1^2)^3\mathcal Q_1^3\mathcal P \mathcal Q}
		\\
		&+\frac{A_{\omega_1}^I}{q_2^2\omega_1^2(q_2^2+\omega_1^2)^3\mathcal P \mathcal Q_1^4}\,
		\operatorname{arcsinh}\frac{\omega_1}{\sqrt{q_2^2}}
		+
		\frac{A_{q_1}^I }{q_2^2\omega_1(\sigma^2-1)\mathcal P^2\mathcal Q^2}\,\frac{\operatorname{arccosh}\sigma}{\sqrt{\sigma^2-1}}\\
		&+
		\frac{A_{\omega_1 \omega_2}^I}{\omega_1\omega_2^2\mathcal P^2\mathcal Q^2}\,\log\frac{\omega_1^2}{\omega_2^2}
		+
		\frac{A_{q_1 q_2}^I }{q_1^2q_2^2 \mathcal Q_1^4\mathcal P\mathcal Q^2}\, \log\frac{q_1^2}{q_2^2} 
		\end{split}
\end{align}
\end{widetext}
with
\begin{align}\label{}
	\begin{split}
	\mathcal P &= -\omega_1^2+2\omega_1\omega_2\sigma-\omega_2^2\,,  \\
	 \mathcal Q &= (q_1^2)^2\omega_1^2-2q_1^2q_2^2\omega_1\omega_2\sigma+(q_2^2)^2\omega_2^2
	\end{split}
\end{align}
marking the appearance of additional spurious poles. The other cut, $c_2$, is again obtained by interchanging the labels 1 and 2 everywhere. Note that, considering $c_1+c_2$ directly would lead to the appearance of both $1/\mathcal Q_1^4$ and $1/\mathcal Q_2^4$ in the rational prefactor in front of $\log(q_1^2/q_2^2)$.

Let us recall that, in terms of the graviton frequency $\omega$ defined in the center-of-mass frame (see Section~\ref{sec:CM} for more details), the following identity holds,
\begin{equation}\label{m1omega1m2omega2}
	m_1 \omega_1 + m_2 \omega_2 = E \omega \,,
\end{equation}
where $E$ is the total energy of the incoming state in that frame.
It is also natural and convenient to define the following dimensionless variable, proportional to the frequency, that does not scale in the PN limit (not to be confused with the retarded time, which we denote by $U$),
\begin{equation}\label{eq:uKT}
	u = \frac{\omega b}{\sqrt{\sigma^2-1}}\,.
\end{equation}

\section{Infrared divergences revisited}
\label{sec:IRdiv}

The standard way of treating the IR divergent piece appearing in the Compton cuts (see \eqref{eq:c1}) is to exponentiate it according to
\cite{Weinberg:1965nx}
\begin{align}\label{eq:expIRdiv}
	\begin{split}
	W^\text{eik}  &= e^{-\frac{i}{\epsilon}\,G E \omega}\left[\mathcal A_0 + \mathcal B_1 + \frac{i}{2}\left(c_1+c_2\right)^\text{reg}\right] \\
	&=
	e^{-\frac{i}{\epsilon}\,G E \omega}
	W^\text{reg}\,,
	\end{split}
\end{align}
where
\begin{equation}\label{eq:c1regDEF}
	\frac{i}{2}\,c_1^{\text{reg}} 
	=
	\frac{i}{2}\,c_1 + \frac{i}{\epsilon}\,G m_1\omega_1 \mathcal A_0
\end{equation}
and similarly for $c_2$.
Note that the $c_1^\text{reg}$ defined in this way is not only IR finite, thanks to the cancellation of the pole $-\frac{i}{\epsilon}\,G m_1 \omega_1\left[\mathcal A_0\right]_{D=4}$ appearing in the first term on the right hand side of \eqref{eq:c1},
but it also has a slightly different finite part compared to the one defined by \eqref{eq:c1}, \eqref{eq:calMm1m2}. This is because $\mathcal A_0$ in \eqref{eq:c1regDEF} is the $D$-dimensional tree-level amplitude, which also contains $\mathcal O(\epsilon)$ terms.\footnote{These arise from the subtraction of the trace, i.e.~of dilaton exchanges. Thus, this subtlety is absent in $\mathcal{N}=8$ supergravity.} The resulting (finite) $\epsilon/\epsilon$ terms thus must be \emph{added} to $\frac{i}{2}\,c_1$ (in particular, they modify $A_{\text{rat}}^I$),\footnote{By the same logic, we expect that the appropriate subtraction of the two-loop kernel will require the knowledge of the $\mathcal O(\epsilon)$ accurate one-loop one, which also involves the full pentagon topology.} leading to
\begin{equation}\label{}
	\frac{i}{2}\,c_1^{\text{reg}}
	= 2i G m_1 \omega_1 \log\frac{\omega_1}{\mu_\text{IR}}
	\mathcal A_0
	+
	i  m_1^3 m_2^2 
	\mathcal M^{m_1^3m_2^2,\text{reg}}\,.
\end{equation}
This is reminiscent of analogous $\epsilon/\epsilon$ contributions featuring in higher-order PN contributions to the MPM-PN approach \cite{Bernard:2017bvn,Blanchet:2023bwj,Blanchet:2023sbv}.
After this step, the divergence in \eqref{eq:expIRdiv} can be renormalized by redefining the origin of retarded time \cite{Goldberger:2009qd,Porto:2012as}, 
\begin{equation}\label{eq:subU}
	U \mapsto U - \frac{1}{\epsilon}\,G E\,,
\end{equation}
arriving at the following well defined expression
\begin{equation}\label{}
	h_{\mu\nu}(x) \sim \frac{4G}{\kappa r}\int_0^\infty
	e^{-i\omega U} \tilde{W}_{\mu\nu}^\text{reg} (\omega n)\,
	\frac{d\omega}{2\pi}+(\text{c.c.})\,,
\end{equation} 
which provides the properly subtracted version of \eqref{eq:naivh}.
Note that performing further $\mathcal O(G)$ finite shifts of the retarded time in \eqref{eq:subU} is equivalent to adjusting the arbitrary scale $\mu_\text{IR}$.
In the following, we will be concerned with the expansion of the spectral waveform $\frac{1}{\kappa}\,\tilde W^{\text{reg}}$ in the PN limit. As we shall see, taking into account the further finite difference in \eqref{eq:c1regDEF} will be crucial in order to obtain agreement with MPM prediction. On the contrary, one should not worry about additional factors of $\epsilon$ introduced by the $2-2\epsilon$ dimensional measure in the  impact-parameter Fourier transform \eqref{eq:FT5}, which should be applied \emph{after} the momentum-space exponentiation (i.e., on the square bracket in \eqref{eq:expIRdiv}).

For later convenience, let us use the identity
\begin{align}\label{}
	&m_1 \omega_1 \log\frac{\omega_1}{\mu_\text{IR}} + m_2 \omega_2 \log\frac{\omega_2}{\mu_\text{IR}} 
	 \nonumber  \\
	&=E \omega \log\frac{\omega}{\mu_{\text{IR}}} + 
	\omega
	\left( 
	m_1 \alpha_1 \log\alpha_1 + m_2 \alpha_2 \log\alpha_2
	\right),
\end{align}
with $\omega_1= \omega \alpha_1$, $\omega_2= \omega \alpha_2$,
to isolate the running logarithm appearing in the (regulated) Compton cuts,
\begin{equation}\label{eq:Creg}
	\frac{i}{2} (c_1+c_2)^{\text{reg}}
	= 2i G E\omega \log\frac{\omega}{\mu_\text{IR}}\,\mathcal A_0 + C^{\text{reg}}
\end{equation}
with
\begin{equation}\label{eq:CregExpl}
	 \begin{split}
C^{\text{reg}}
&= 2iG \omega  m_1 \alpha_1\log\alpha_1\,\mathcal A_0 + i \, m_1^3m_2^2 \, \mathcal M^{m_1^3m_2^2,\text{reg}} \\
&+ (1\leftrightarrow2)\,.
	 \end{split}
\end{equation}

\section{Post-Newtonian Expansion of the Amplitude-Based Waveform}
\label{sec:PNampl}

In this section, we discuss the expansion of the waveform obtained from the regulated amplitude kernel $W^\text{reg}$ in \eqref{eq:expIRdiv} in the PN limit. We start from the real part of the waveform kernel and then turn to its imaginary part.
We will proceed by taking the PN limit already from the outset, in momentum space, discussing below the challenges that arise in this step.
Once the expansion is performed, all $q$-dependent spurious poles manifestly cancel out. The remaining one, arising from the expansion of  $\mathcal{P}$, can then be  easily removed by taking the polynomial remainder with respect to the Gram determinant, which implements the four-dimensional identity \eqref{eq:gram}.

\subsection{Kinematic conventions}
\label{sec:CM}
We will consider the PN limit in the center-of-mass frame. We thus introduce the four-velocity $t^\mu$ of the center-of-mass frame and the unit vector $e^\mu$ aligned with the ``eikonal'' particles' momenta in the center-of-mass frame, 
\begin{equation}\label{}
	t^\mu = (1,0,0,0)\,,\qquad e^\mu = (0,0,1,0)\,,
\end{equation}
so that
\begin{subequations}
	\begin{align}
		\tilde u_1^\mu &= \frac{E_1}{m_1} \,t^\mu + \frac{p}{m_1}\,e^\mu
		=
		\frac{1}{m_1}\left(
		E_1,0,+p,0
		\right)
		,\\
		\tilde u_2^\mu & =  \frac{E_2}{m_2} \,t^\mu - \frac{p}{m_2}\,e^\mu
		=
		\frac{1}{m_2}\left(
		E_2,0,-p,0
		\right),
	\end{align}
\end{subequations}
where $E_{1,2}$ denote the particles' energies,
\begin{align}\label{}
	E_1 &= \frac{m_1(m_1+m_2\sigma)}{\sqrt{m_1^2+2m_1m_2\sigma+m_2^2}}\,,  \nonumber \\ 
	E_2 &= \frac{m_2(m_2+m_1\sigma)}{\sqrt{m_1^2+2m_1m_2\sigma+m_2^2}}
\end{align}
and
\begin{align}\label{}
	E &= E_1 + E_2  = \sqrt{m_1^2+2m_1m_2\sigma+m_2^2}\,, \nonumber \\
	p &= \frac{m_1m_2\sqrt{\sigma^2-1}}{\sqrt{m_1^2+2m_1m_2\sigma+m_2^2}}\,.
\end{align}
In terms of the ``eikonal'' impact parameter, we define
\begin{equation}\label{key}
	b_e^\mu = (0,b,0,0)\,.
\end{equation}
Moreover, we fix the translation frame by letting
\begin{align}
	b_1^\mu &= +\frac{E_2}{E}\, b_e^\mu =
	+\frac{E_2}{E} (0,b,0,0)\,, \nonumber \\
	b_2^\mu &= -\frac{E_1}{E}\, b_e^\mu =
	-\frac{E_1}{E} (0,b,0,0)\,,
\end{align}
so that $E_1 b_1^\mu + E_2 b_2^\mu= 0$.
Working in this frame, we solve the delta functions in \eqref{eq:FT5} as follows
\begin{equation}\label{eq:FT5CM}
	\begin{split}
		\tilde W_{\mu\nu}(k)
		&=  \frac{1}{4 m_1 m_2 \sqrt{\sigma^2-1}}\,e^{-ib_2\cdot k}
		\\
		&\times \int \frac{d^{2} q_\perp}{(2\pi)^2}\,  W(q_1,-k-q_1)\, e^{ib\cdot q_\perp}
	\end{split}
\end{equation}
and by letting
\begin{equation}\label{}
	q_1^\mu = q_\perp^\mu - \omega_2\, \frac{\sigma \tilde u_1^\mu-\tilde u_2^\mu}{\sigma^2-1}\,.
\end{equation}
As a result, in the PN limit, thanks to the cancellation of the spurious poles discussed above, all Fourier transforms can be easily evaluated with the help of the general formula
\begin{equation}\label{eq:FTwithomegapinf}
	\begin{split}
		\int \frac{d^{2}q_\perp}{(2\pi)^{2}}
		\left(
		1+\frac{p_\infty^2q_\perp^2}{\omega^2}
		\right)^\nu\!
		e^{ib_e\cdot q_\perp}=
		\frac{2^\nu}{\pi b^2} \frac{K_{1+\nu}\left(u\right)}{\Gamma(-\nu)\,u^{\nu-1}}\,,
	\end{split}
\end{equation}
where $K_\alpha(x)$ denotes the modified Bessel function of the second kind.
We let
\begin{equation}\label{}
	k^\mu = \omega\, n^\mu\,, 
\end{equation}
with $n^\mu$ a null vector such that $-n\cdot t=1$, so that $\omega$ is the frequency as measured in the center-of-mass frame.
It is also useful to introduce the total mass $m$ and the symmetric mass ratio $\nu$ by
\begin{equation}\label{}
	m=m_1+m_2,\qquad
	\nu= \frac{m_1 m_2}{m^2}\,.
\end{equation}

Using this notation, we define the PN expansion by the scaling limit 
\begin{equation}\label{eq:limitPN}
	p_\infty = \sqrt{\sigma^2-1} = \mathcal O(\lambda)\,,\qquad \omega = \mathcal O(\lambda)
\end{equation}
as $\lambda \to 0$. 
In the conventional PN counting, each instance of the Newton constant $G$ increases the PN order by one unit, while each power of $\lambda$ increases it by half a unit. Hence, odd powers of $\lambda$ multiplying a given power of $G$ can lead to the appearance of fractional PN orders.

Let us also record the explicit expressions
\begin{align}
\omega_1&=\frac{\omega}{E} \left(m_1+m_2 \left( \sigma -p_\infty n \cdot e \right) \right), \\
\omega_2&=\frac{\omega}{E} \left(m_2+m_1 \left( \sigma +p_\infty n \cdot e \right) \right)
\end{align}
and remark that, in the expansion, there is an important difference between powers of $\lambda$ that appear accompanied by an extra $n$ and ``bare'' ones. The latter change the scaling of a given multipole, while the former produce contributions to different multipoles. Using this property, it is easy to see from the outset that, for instance, $\mathcal B_{1E}$ only contributes to \emph{integer} PN corrections to a given multipole, while $C^{\text{reg}}$ only gives rise to \emph{half-odd} corrections, that is, time-reversal odd terms for each multipole. We refer to \cite{Georgoudis:2023eke} for further details on this counting.

In order to simplify the PN expansion, it is convenient to first perform it for (say) the $m_1^3m_2^2$ mass structure, and then exploit $1\leftrightarrow2$ symmetry to obtain the expansion of the other structure. Note that the action of the interchange $1\leftrightarrow2$ symmetry is not trivial on all variables after the PN expansion, owing to the choice \eqref{eq:FT5CM} which privileges $q_1$ over $q_2$.

\vspace{5pt}

\subsection{The real part of the kernel}

It is straightforward to expand the tree-level amplitude $\mathcal A_0$ up to high orders in the PN limit \eqref{eq:limitPN}. A particularly convenient way to approach this expansion is to first compute the power series of the elementary variables $\sigma$, $\omega_1$, $\omega_2$, $q_1^2$, $q_2^2$, $\varepsilon\cdot u_1$, $\varepsilon\cdot u_2$, $\varepsilon\cdot q_2$ and then substitute them into the full expression. 

After this step, we explicitly performed the Fourier transform to impact-parameter space \eqref{eq:FT5CM} and the multipolar decomposition, which in general takes the form
	\begin{equation}\label{Blanchet66text}
		\begin{split}
			\mathcal H_{ij}^\text{TT} 
			&= 
			\mathcal P_{ijab}(n)\sum_{\ell=2}^\infty\frac{1}{\ell!}
			\Bigg[
			n_{L-2}\, \mathrm{U}_{abL-2}(u)
			\\
			&-\frac{2\ell}{\ell+1}\,n_{cL-2}\epsilon_{cd(a} \, \mathrm{V}_{b)dL-2}(u)
			\Bigg]
		\end{split}
	\end{equation}
as summarized in Appendix~\ref{sec:multipoles}, for the first three orders, $\lambda^{-1}$, $\lambda^0$, $\lambda^1$ and $\lambda^{2}$. Following the standard nomenclature, these contribute to the Newtonian (0PN), 0.5PN, 1PN and 1.5PN orders in the asymptotic waveform.
We collect in the ancillary file all the associated multipoles, which are the leading (Newtonian) order contributions to $\mathrm{U}_{2,3,4,5}$ and $\mathrm{V}_{2,3,4}$, as well as the first subleading (relative 1PN) contribution to $\mathrm{U}_{2,3}$ and $\mathrm{V_2}$.
As already remarked, by counting the ``dressed'' and ``bare'' powers of $\lambda$, it is easy to see that the tree-level contribution to $\mathrm{U}_L$ (resp.~$\mathrm{V}_L$) starts at order $\lambda^{\ell-3}$ (resp.~$\lambda^{\ell-2}$) and only receives corrections weighted by \emph{even} powers of $\lambda$.
Here, we present as an example the LO (Newtonian) quadrupole
\begin{subequations}\label{eq:U2LO}
	\begin{align}
		\mathrm{U}_{11}^{\text{LO}}
		&= -\frac{4 G m^2 \nu}{3 p_\infty}(K_0(u)+3 u K_1(u))\,,\\
		\mathrm{U}_{12}^{\text{LO}}
		&= 	-\frac{4 i G m^2 \nu}{p_\infty}(u K_0(u)+K_1(u))\,,\\
		\mathrm{U}_{22}^{\text{LO}}
		&= \frac{4 G m^2 \nu}{3 p_\infty} (2 K_0(u)+3 u K_1(u))\,,\\
		\mathrm{U}_{33}^{\text{LO}}
		&= -\frac{4 G m^2 \nu K_0(u)}{3 p_\infty}\,.
	\end{align}
\end{subequations}
and its first subleading, 1PN, correction,
\begin{widetext}
\begin{subequations}
	\begin{align}\label{}
		\mathrm{U}_{11}^{\text{NLO}}
		&= \frac{2}{21} G \nu  m^2 p_{\infty } \left(\left(-8 \nu +(19-36 \nu ) u^2+26\right) K_0(u)-7 (4 \nu +3) u K_1(u)\right),\\
		\mathrm{U}_{12}^{\text{NLO}}
		&= 	-\frac{2}{21} i G \nu  m^2 p_{\infty } \left(\left(24 \nu +(36 \nu -19) u^2+69\right) K_1(u)+(18 \nu  u+u) K_0(u)\right),\\
		\mathrm{U}_{22}^{\text{NLO}}
		&= \frac{2}{21} G \nu  m^2 p_{\infty } \left(\left(16 \nu +(36 \nu -19) u^2-52\right) K_0(u)+2 (22 \nu -19) u K_1(u)\right),\\
		\mathrm{U}_{33}^{\text{NLO}}
		&= \frac{2}{21} G \nu  m^2 p_{\infty } ((26-8 \nu ) K_0(u)+(59-16 \nu ) u K_1(u))\,.
	\end{align}
\end{subequations}
\end{widetext}  
As expected, these quantities are exponentially suppressed for $u\gg1$, i.e.~for wavelengths much shorter than the characteristic length $b/p_\infty$ of the binary.

Let us now turn to the real part of the one-loop kernel, $\mathcal B_1^{\mu\nu}$, starting from its ``odd'' part $\mathcal B_{1O}^{\mu\nu}$ given by \eqref{eq:B1odd}. 
Since $\tilde{\mathcal{A}}_0$ scales like $\lambda^{-1}$ to leading PN order, clearly $\tilde{\mathcal{B}}_{1O}^{(i)}$ scales like $\lambda^{-3}$ and $\tilde{\mathcal{B}}_{1O}^{(h)}$ like $\lambda^0$. In view of the overall power of $G^2$, the PN expansion of $\mathcal{B}_{1O}$ thus starts at Newtonian order.
Moreover, given that the prefactors in \eqref{eq:B1odd} are angle-independent, the expansion of $\mathcal{B}_{1O}$ and the associated multipoles $\mathrm{U}_{O L}$, $\mathrm{V}_{O L}$ are trivially related to those of the tree-level waveform. For instance, the LO contribution of each one-loop multipole comes from $\mathcal B_{1O}^{(i)}$ in \eqref{eq:B1odd(i)} and is fixed by the corresponding tree-level ones by the simple relation \cite{Bini:2023fiz}
\begin{equation}\label{LONLOrelationPi}
	\mathrm{U}_{O L}^{\text{LO}} = 
	\frac{\pi G m u}{2b\,p_\infty^2}\, \mathrm{U}^{\text{LO}}_{L}
	\,,
	\qquad
	\mathrm{V}_{O L}^{\text{LO}} = 
	\frac{\pi G m u}{2b\,p_\infty^2}\, \mathrm{V}^{\text{LO}}_{L}
	\,.
\end{equation}
Indeed, the overall power of $\frac{Gm}{bp_\infty^2}$ is dimensionless and small in the PM limit, but, in view of the power counting rule detailed below \eqref{eq:limitPN}, it does not increase the PN order.
The subleading corrections are instead given by linear combinations of the leading and subleading tree-level multipoles as dictated by \eqref{eq:B1odd}. 
For instance, the 1PN correction to the quadrupole reads
\begin{widetext}
\begin{subequations}
	\begin{align}
		\mathrm{U}_{O11}^\text{NLO}
		&=
		\frac{\pi  G^2 \nu  m^3 u \left(\left(-15 \nu +(19-36 \nu ) u^2+47\right) K_0(u)+7 (6-7 \nu ) u K_1(u)\right)}{21 b p_{\infty }}\,,
		\\
		\mathrm{U}_{O12}^\text{NLO}
		&=
		-\frac{i \pi  G^2 \nu  m^3 u \left(\left(45 \nu +(36 \nu -19) u^2+6\right) K_1(u)+(39 \nu -62) u K_0(u)\right)}{21 b p_{\infty }}\,,
		\\
		\mathrm{U}_{O22}^\text{NLO}
		&=
		\frac{\pi  G^2 \nu  m^3 u \left(\left(30 \nu +(36 \nu -19) u^2-94\right) K_0(u)+(65 \nu -101) u K_1(u)\right)}{21 b p_{\infty }}\,,
		\\
		\mathrm{U}_{O33}^\text{NLO}
		&=\frac{\pi  G^2 \nu  m^3 u ((47-15 \nu ) K_0(u)+(59-16 \nu ) u K_1(u))}{21 b p_{\infty }}\,,
	\end{align}
\end{subequations}
\end{widetext}
which is given by the  
expansion of the prefactor in \eqref{eq:B1odd(i)},
\begin{equation}\label{}
	\mathrm{U}_{Oij}^\text{NLO}
	=
	\frac{\pi G m u}{2b\,p_\infty^2} \mathrm{U}_{ij}^\text{NLO}
	+
	\frac{G m \pi u}{4b}(\nu-3) \, \mathrm{U}_{ij}^\text{LO}\,.
\end{equation}
The fact that the structure above generalises, via the simple formula \eqref{eq:B1odd(i)}, to all PN orders is rather remarkable from the MPM perspective. The amplitude result implies that the integer PN contributions involving $K_0(u)$ and $K_1(u)$ in the $\mathcal O(G^2)$ multipoles are captured by $\mathcal B_{1O}^{(i)}$.
Instead, the PN expansion of $\mathcal B_{1O}^{(h)}$ in \eqref{eq:B1odd(h)} only involves relative half-odd PN corrections, and, as we shall see in Section~\ref{sec:MPM}, its origin in the MPM formalism can be traced back to the tail formula.

The leading-order contribution to $\mathcal B_{1E}^{\mu\nu}$ in \eqref{eq:B1even} arises at order $\lambda^{-1}$, that is 1PN, and is captured by a correction $\mathrm{U}_{E ij}$ to the quadrupole whose nonvanishing components we find to be given by
\begin{subequations}\label{eq:UEleading}
	\begin{align}
		\mathrm{U}_{E 11}^\text{LO}
		&= -\mathrm{U}_{E 22} = - \frac{6 \pi G^2 m^3 \nu}{b p_\infty}\left(1+u\right) e^{-u}\,,
		\\
		\mathrm{U}_{E 12}^\text{LO}
		&=-\frac{6 i \pi G^2 m^3 \nu}{b p_\infty}\left(\frac{1}{u}+ 1 +u  \right)e^{-u}\,,
	\end{align}
\end{subequations}
in agreement with \cite{Bini:2023fiz}.
We extended this result by including three more orders in $\lambda$ and extracted the multipoles that determine the PN expansion of $\mathcal B_{1E}^{\mu\nu}$ up to NNNLO, reaching absolute 2.5PN order, and we collect them in the ancillary files. For instance, the 2PN contribution to the quadrupole (that is, the relative 1PN correction  to \eqref{eq:UEleading}) is given by
\begin{widetext}
\begin{subequations}
	\begin{align}\label{eq:56a}
		\mathrm{U}_{E11}^\text{NLO}	&= 
		-\frac{\pi  G^2 m^3 \nu  p_{\infty }}{14 b}
		\,
		(52 \nu +u (52 \nu +(78 \nu -61) u+27)+27)\, e^{-u}\,,
		\\ \label{eq:56b}
		\mathrm{U}_{E12}^\text{NLO} &=
		-\frac{i \pi  G^2 m^3 \nu   p_{\infty } }{14 b u}
		\,
		(48 \nu +u (48 \nu +u (66 \nu +(78 \nu -61) u-8)+75)+75)\,e^{-u}\,,
		\\ \label{eq:56c}
		\mathrm{U}_{E22}^\text{NLO} &=
		\frac{\pi  G^2 m^3 \nu  p_{\infty }}{14 b}\,
		(80 \nu +u (80 \nu +(78 \nu -61) u-43)-43)\,e^{-u}\,,
		\\ \label{eq:56d}
		\mathrm{U}_{E33}^\text{NLO} &=
		-\frac{\pi  G^2 m^3 \nu  p_{\infty }}{b}\, (2 \nu -5)  (u+1)\,e^{-u}\,.
	\end{align}
\end{subequations}
\end{widetext}
As a cross-check, we independently derived $\mathrm{U}_{E33}$ from MPM methods finding perfect agreement with \eqref{eq:56d} (see Section \ref{sec:MPM} below). 

Corrections of this type arise from several contributions due to the PN deformation of the quasi-Keplerian representation of the trajectories in the MPM formalism. On the other hand, they all arise from the Taylor expansion as $\lambda \to 0$ of a single object, $\mathcal B_{1E}$ in the amplitude-based formalism. The main difficulty in performing this series expansion to high orders is the presence of spurious poles in the PM expression \eqref{eq:B1even}. Such apparent singularities in $\lambda$ eventually cancel out only thanks to cross-cancellations between the two square roots. This however requires expanding the numerators up to high powers of $\lambda$ in order to obtain a reliable result. For the expansion of $\mathcal B_{1E}$, we found that \textit{Mathematica} is nevertheless able to produce the desired expansion and to analytically simplify the result up to NNLO. To extract the NNNLO, we employed instead the method of analytic reconstruction discussed below for the Compton cuts $\frac{i}{2}(c_1+c_2)$, in which as we shall see this problem is even more acute. 

Let us now perform a first check of the results obtained so far. Starting from the multipoles extracted from $\mathcal A_0$, $\mathcal B_{1O}$ and $\mathcal B_{1E}$, substituting them into the energy-flux formula~\eqref{eq:dEdo} and integrating over $\omega$, we obtain the following expression for the total emitted energy $E_\text{rad}$,
\begin{align}\label{eq:Erad}
		&E_\text{rad}/(m \nu^2)
		\\
		\nonumber
		&=\frac{G^3m^3}{b^3}\,\pi p_\infty
		\left[
		\frac{37}{15}
		+
		\left(\frac{1357}{840}-\frac{37 \nu }{30}\right) p_\infty^2
		+\mathcal O(p_\infty^4)
		\right]
		\\
		\nonumber
		&+
		\frac{G^4m^4}{b^4p_\infty}\left[
		\frac{1568}{45}
		+
		\left(\frac{18608}{525}-\frac{1136}{45}\nu\right) p_\infty^2
		+\mathcal O(p_\infty^4)
		\right]
		\\
		\nonumber
		&+
		\frac{G^4m^4}{b^4}p_\infty^2\left[
		\frac{3136}{45}
		+
		\left(\frac{1216}{105}-\frac{2272}{45}\nu\right) p_\infty^2
		+\mathcal O(p_\infty^4)
		\right]
		\\
		\nonumber
		&+\mathcal O(G^5)\,.
\end{align}
The first two lines on the right-hand side of \eqref{eq:Erad} arise from $\mathcal A_0$, $\mathcal B_{1O}^{(i)}$, $\mathcal B_{1E}$ and reproduce the known $\mathcal O(G^3)$ and $\mathcal O(G^4)$ of the total emitted energy for the scattering to leading and subleading order in the velocity expansion \cite[Eqs.~(C11), (C12)]{Bini:2021gat} (we recall that, setting $G=m=1$, the parameter $j$ employed there reads $j=bp_\infty/\sqrt{1+2\nu(\sigma-1)}$ in our notation). The third line on the right-hand side of \eqref{eq:Erad} arises instead from $\mathcal A_0$, $\mathcal B_{1O}^{(h)}$ and matches the 1.5PN and 2.5PN terms in the $\mathcal O(G^4)$ emitted energy \cite{Dlapa:2022lmu,Bini:2022enm}.
The multipoles arising from $C^\text{reg}$ discussed in the next section give a vanishing contribution to the energy flux when inserted in \eqref{eq:dEdo}, due to the different powers of ``$i$'' compared to the tree level ones.

\subsection{The $C$-channel cuts}

The portion of the one-loop kernel that multiplies the running logarithm in \eqref{eq:Creg} associated to the tail effect is trivially related to the tree-level waveform, up to the angle-independent prefactor
\begin{equation}\label{}
	2iG E\omega=2i u p_\infty\sqrt{1+2\nu(\sigma-1)}\,\frac{Gm}{b}\,.
\end{equation}
Hence, the corresponding multipoles are easily determined by the tree-level ones. All leading-order multipoles multiplying the $\log(\omega/\mu_\text{IR})$ are given by
\begin{equation}\label{}	
	\mathrm{U}_{{\text{IR}}\, L}^\text{LO} = 2i u p_\infty\,\frac{Gm}{b}\,\mathrm{U}_{L}^\text{LO}\,,\quad
	\mathrm{V}_{{\text{IR}}\, L}^\text{LO} = 2i u p_\infty\,\frac{Gm}{b}\,\mathrm{V}_{L}^\text{LO}\,,
\end{equation}
while for instance
\begin{equation}\label{}
	\mathrm{U}_{{\text{IR}}\, ij}^\text{NLO} = 
	\left(
	2i u p_\infty\,\mathrm{U}_{L}^\text{NLO}
	+ i u \nu p_\infty^3 \mathrm{U}_{L}^\text{LO}
	\right)
	\frac{Gm}{b}
	\,.
\end{equation}

The PN expansion of the remaining part $C^\text{reg}$  of the Compton cuts in \eqref{eq:Creg} is instead less straightforward, and we devote the rest of this section to its discussion. It starts at 1.5PN order compared with the one-loop Newtonian terms, that is, $\mathcal O(\lambda^0)$. We have extracted its multipole decomposition up to and including $\mathcal O(\lambda^3)$, thus reaching 3PN precision. We find perfect agreement with the MPM predictions upon taking into account the effect of the BMS supertranslation that maps the canonical frame to the intrinsic frame, which amounts to transforming \eqref{eq:CregExpl} as follows, 
\begin{equation}\label{eq:ST}
	\begin{split}
C^\text{reg} 
&\mapsto C^{\text{reg}}	\\
&+
2iG \omega  \left(m_1 \alpha_1\log\alpha_1+m_2 \alpha_2\log\alpha_2\right) \mathcal A_0\,.
	\end{split}
\end{equation}
The complete action of the supertranslation (see e.g.~\cite[Eq.~(5.24)]{Georgoudis:2023eke}) would also contain a static term in time domain, i.e.~a contribution localized at $\omega=0$ in frequency domain, which we disregard because we focus on the $\omega>0$ portion of the spectrum.
The transformation \eqref{eq:ST} simply amounts to multiplying the first term on the right-hand side of \eqref{eq:CregExpl} by two. 

We highlight this in the ancillary file by the parameter \texttt{tail} (which should be set to 2 in order to take \eqref{eq:ST} into account).
We also highlight nontrivial $\epsilon/\epsilon$ terms arising from the subtraction of infrared divergences \eqref{eq:c1regDEF} by flagging them with a parameter \texttt{extraeps} (which should be set to 1).

In this way, for instance, for the quadrupole we find
\begin{subequations}
	\begin{align}
		\mathrm{U}_{C11} &=\frac{22 i G^2 m^3 \nu  u (K_0(u)+3 u K_1(u))}{9 b}\,,\\
		\mathrm{U}_{C12} &=-\frac{22 G^2 m^3 \nu  u (u K_0(u)+K_1(u))}{3 b}\,,\\
		\mathrm{U}_{C22} &=-\frac{22 i G^2 m^3 \nu  u (2 K_0(u)+3 u K_1(u))}{9 b}\,,\\
		\mathrm{U}_{C33} &=\frac{22 i G^2 m^3 \nu  u K_0(u)}{9 b}
	\end{align}
\end{subequations}
to leading order, and 
\begin{widetext}
\begin{subequations}\label{eq:63}
	\begin{align}
		\mathrm{U}_{C11}^{\text{NLO}} &=\frac{i G^2 m^3 \nu  u p_{\infty }^2 \left(5 \left(141 \nu +(984 \nu -209) u^2-286\right) K_0(u)+7 (373 \nu +165) u K_1(u)\right)}{315 b}\,,\\
		\mathrm{U}_{C12}^{\text{NLO}} &=-\frac{G^2 m^3 \nu  u p_{\infty }^2 \left(\left(2043 \nu +5 (984 \nu -209) u^2+3795\right) K_1(u)+(1461 \nu +55) u K_0(u)\right)}{315 b}\,,\\
		\mathrm{U}_{C22}^{\text{NLO}} &=-\frac{i G^2 m^3 \nu  u p_{\infty }^2 \left(5 \left(282 \nu +(984 \nu -209) u^2-572\right) K_0(u)+(5231 \nu -2090) u K_1(u)\right)}{315 b}\,,\\
		\mathrm{U}_{C33}^{\text{NLO}} &=\frac{i G^2 m^3 \nu  u p_{\infty }^2 ((141 \nu -286) K_0(u)+(524 \nu -649) u K_1(u))}{63 b}
	\end{align}
\end{subequations}
\end{widetext}
for the relative 1PN/absolute 2.5PN correction. 
We list the remaining multipoles in the ancillary files.

For the Compton cuts, the high degree of the spurious singularities and their intricate cross-cancellations among logarithms (+ rational part) made it impractical for us to simply expand their expression to the desired order in $\lambda$ and analytically simplify the result.
However, when focusing on (say) $c_1$, we found it very quick to evaluate its series expansion on numerical points. 
We exploited this, together with the knowledge of the possible poles dictated by the PM expression \eqref{eq:c1}, to obtain the desired expansion in the following way. In momentum-space, the expressions are all rational functions after the PN expansion. Therefore, making a sufficiently general polynomial  ansatz for the numerators, it is then possible to fix it by sampling enough numerical points, which we can efficiently do. We need 180 points to fix the coefficient of $(\varepsilon\cdot e)^2$ in $c_1$ at 3PN for a given mass ratio. 

As we shall see, the Compton cuts are sensitive to various kinds of physical effects.
First, the $\mathcal O(\nu)$ part of the associated multipoles is entirely fixed in the MPM approach by the tail formula.
Second, the $\mathcal O(\nu^2)$ terms receive several contributions from nonlinear relations between lower order multipoles,
dictated by the nonlinearities of gravity, by the matching of the near-zone and far-zone multipoles, and by radiation-reaction effects.

We conclude this section by reproducing some of the results of~\cite{Bini:2021gat,Bini:2022enm} for the the radiated spatial momentum. By integrating over $\omega$ the expression of the linear momentum flux~\eqref{eq:dPdo} in terms of the multipoles, we obtain
\begin{align}\label{eq:Prad}
	&P^\mu_\text{rad}/(m \nu^2 \sqrt{1-4\nu})
	\\
	\nonumber
	&=\frac{G^3m^3}{b^3} \,
	\pi
	\left[
	-\frac{37}{30}
	p^2_\infty
	+
	\left(
	\frac{37}{60}\nu
	-
	\frac{839}{1680}
	\right)
	p_\infty^4
	+\mathcal O(p_\infty^6)
	\right]e^\mu
	\\
	\nonumber
	&+
	\frac{G^4m^4}{b^4}\left[
	-\frac{64}{3}
	+
	\left(
	\frac{32}{3}\nu
	-
	\frac{1664}{175}
	\right)
	p_\infty^2
	+\mathcal O(p_\infty^4)
	\right]
	e^\mu
	\\
	\nonumber
	&+
	\frac{G^4m^4}{b^4}\,p_\infty^3
	\left[
	\left(\frac{1491}{400} - \frac{26757}{5600}p_\infty^2\right) \pi\,\frac{b_e^\mu}{b}
	\right.
	\\
	\nonumber
	&+
	\left.
	\left(
	-\frac{128}{3} 
	+\left(
	\frac{64}{3}\nu- \frac{192}{75}
	\right)p_\infty^2
	\right) e^\mu
	+\mathcal O(p_\infty^4)
	\right]
	\\
	\nonumber
	&+\mathcal O(G^5)\,.
\end{align}
The first two lines on the right-hand side of \eqref{eq:Prad} arise from $\mathcal A_0$, $\mathcal B_{1O}^{(i)}$, $\mathcal B_{1E}$, and reproduce the known $\mathcal O(G^3)$ and $\mathcal O(G^4)$ of the total emitted spatial momentum for the scattering to leading and subleading order in the velocity expansion \cite[Eqs.~(G7), (G8)]{Bini:2021gat}. The third and fourth lines on the right-hand side of \eqref{eq:Prad} arise instead from $\mathcal A_0$, $\mathcal B_{1O}^{(h)}$, $C^\text{reg}$ and match the $\mathcal O(G^4)$ total emitted spatial momentum up to 2.5PN \cite{Bini:2021gat,Bini:2022enm}.
More precisely, the component of \eqref{eq:Prad} along $b_e^\mu$ arises entirely from the interference terms between $\mathcal A_0$ and $C^{\text{reg}}$, providing interesting cross-checks on the latter. The term proportional to $\frac{1491}{400}$ follows entirely from the interference with the tail contribution to the multipoles, while the next order in the same line is also sensitive to nonlinear and radiation-reaction contributions to $C^\text{reg}$, as well as to the $\epsilon/\epsilon$ terms induced by \eqref{eq:c1regDEF}. We recall that $b_e\cdot P_\text{rad}/b$ is instrumental in deriving the so-called $(\text{radiation-reaction})^2$ contributions to the $\mathcal O( G^4 )$ impulse \cite{Dlapa:2022lmu} as shown in \cite{Bini:2022enm}.

\section{Multipolar post-Minkowskian Waveform}
\label{sec:MPM}

In this section, we briefly summarize the ingredients needed to derive the post-Newtonian expansion of the NLO PM scattering waveform following the classical GR literature~\cite{Blanchet:2013haa}. The idea is to relate the MPM expansion to the dynamics of the source ({\rm i.e.} the binary in our case) by matching the far-zone result
with the near-zone gravitational field.\footnote{In practice the matching is done in two steps, between a far and an intermediate zone, and then between this intermediate and the near zone~\cite{Blanchet:2013haa}. The formulae we use in this section combine these two steps.} The latter is parametrized in terms of STF tensors $I_L$, $J_L$, whose role is reminiscent\footnote{{We use italic symbols to indicate the time-domain  multipoles and roman symbols for their Fourier transform in frequency domain as in~\eqref{eq:UitUrm}.}} of ${U}_L$ and ${V}_L$, and by some auxiliary moments, $W_L$, $X_L$, $Y_L$ and $Z_L$, which encode a linearized gauge transformation. Explicit expressions for the near-zone moments are obtained by solving iteratively Einstein's equations using the stress-tensor of the binary. In summary, thanks to the matching procedure mentioned above, one can write (perturbatively) ${U}_L$, ${V}_L$ in terms the near zone moments and the latter in terms of the binary trajectory thus obtaining an explicit PN solution for the waveform. In this approach, it is convenient to work in the time domain, so, in order to compare with the radiative multipoles $\mathrm{U}_L(u)$, $\mathrm{V}_L(u)$ obtained from scattering amplitudes, we will need to take the Fourier transform (let us recall that $u$ is the rescaled frequency defined in \eqref{eq:uKT})
\begin{equation}\label{eq:UitUrm}
	\mathrm{U}_L(u) = \int_{-\infty}^{+\infty} U_{L}(t)\,e^{i\omega t}\,{dt}\,,
\end{equation}
and similarly for $\mathrm{V}_L$. In our case, it was always possible to perform this step by using
\begin{equation}\label{eq:FTBesselPN}
 	\int_{-\infty}^{+\infty} \frac{e^{i\omega t} \, dt}{[1+(p_\infty t/b)^2]^{\alpha}}
 	=
 	\frac{2^{\frac{3}{2}-\alpha}\sqrt{\pi}}{\omega}\frac{K_{\alpha-\frac{1}{2}}(u)}{u^{-\alpha-\frac{1}{2}}\Gamma(\alpha)}
\end{equation}
and its derivatives. In the equation above, $K_a(b)$ is the usual modified Bessel function and we recall that $u=\frac{\omega b}{p_\infty}$ as in \eqref{eq:uKT}. 

In this work we will consider a further truncation in $G$ of the PN expansion, in order to compare with the PM results. 
However, at low PN orders, it is also possible to take a different approach and perform the Fourier transform to frequency domain by keeping the dependence on $Gm/(b p_\infty^2)$ exact. In that case, one obtains Bessel functions $K_a(b)$, where the order $a$ deviates from an integer value by a shift proportional to $Gm\omega/p_\infty^3$~\cite{Bini:2021jmj}. 

It is convenient to separate three types of contributions to the MPM multipoles: a first part, which is local in time and depends on the conservative dynamics of the binary, a hereditary part, which at time $U$ depends on the whole past history of the binary, and a radiation-reaction part, which takes into account the dissipative effects on the binary trajectory. According to this splitting, we write for instance
\begin{equation}
  \label{eq:Udec}
  {U}_L = {U}^{\rm loc}_L + {U}^{\rm hered}_L + {U}^{\rm RR}_L \,.
\end{equation}

\subsection{Binary trajectory and near zone multipoles}

It is convenient to describe the relative distance between the two components of the binary in terms of polar coordinates on the scattering plane $x= r \cos\phi$, $y=r \sin\phi$ and then use the following quasi-Keplerian parametrization of the trajectory (see~\cite{Cho:2018upo,Bini:2021jmj} and references therein), 
\begin{subequations}\label{eq:QKp}
	\begin{align} \label{eq:QKpa}
r &= Gm a_r \left(e_r \cosh v-1\right),\\ \label{eq:QKpb}
          \tilde{\phi} & = 2 \arctan \left( \sqrt{\frac{e_\phi+1}{e_\phi-1}}\,\tanh\frac{v}{2}\right) , \\ \label{eq:QKpc}
n t &= Gm \left(e_t \sinh v-v + f_t \tilde\phi + g_t \sin\tilde\phi\right) + \mathcal O(c^{-6}),\\ \label{eq:QKd}
\phi &= K \left[ \tilde\phi + f_\phi \sin(2\tilde\phi) + g_\phi \sin(3\tilde\phi) + \mathcal O(c^{-6})
\right].
	\end{align}
\end{subequations}
At leading PN order, the eccentricities are all equal, $e_\phi \approx e_t \approx e_r \approx b p_\infty^2/(Gm)$, and we have $K \approx 1$, $a_r \approx p_\infty^{-2}$, $n\approx p_\infty^{-3}$, while $f_{t,\phi}$ and $g_{t,\phi}$ start at $\mathcal O(c^{-4})$. We refer to~\cite{Cho:2018upo,Bini:2021jmj} for the explicit expressions up to 3PN of all parameters defining above, but here let us just quote $f_t\approx 3(5-2\nu) Gm p_\infty^2/b$ which will be useful later. We follow here the conventional notation according to which higher PN orders are highlighted by inverse powers of $c$, the speed of light. Notice that, in the leading PM trajectory, one can easily solve the relation between $t$ and $v$, obtaining the straight-line motion $x = b$,  $y = p_\infty t$.

The near zone field is then written in terms of the binary trajectory and, since this is a gauge-dependent quantity, some care is needed. It is convenient to use the so-called modified-harmonic coordinates where the instantaneous part of the quadrupole moment $I_2$ is given by~\cite{Mishra:2015bqa}
\begin{equation}\label{I2loc}
I^{\rm inst}_{ij} = m \nu\! \left[ A_1  x_{\langle i} x_{j\rangle} + A_2 \frac{r \dot{r}}{c^2}  x_{\langle i} \dot{x}_{j\rangle} + A_3 \frac{r^2}{c^2}  \dot{x}_{\langle i} \dot{x}_{j\rangle}\right]
\end{equation}
where we use the dot to indicate the time derivative, for instance $\dot{r} = \frac{dr}{dt}$. The $A_i$'s are themselves functions of the mass ratio and the trajectories known in a PN expansion starting at order $c^0$, for instance $A_1=1+\mathcal O(c^{-2})$ (see Eq.(3.20) of~\cite{Mishra:2015bqa} for the expressions up to 3PN). While the PN expansion of~\eqref{I2loc} contains only even powers of $1/c$, thus leading to integer relative PN corrections, starting at order $1/c^5$ there are new contributions related to use of retarded propagators, as discussed in \cite{Blanchet:1996wx}. For the near zone quadrupole they are captured by the following result at $\mathcal O(c^{-5})$ precision
\begin{equation}\label{I2ret}
  I^{\rm ret}_{ij} = 
      \frac{G^2\nu^2}{r^2 c^5}\,m^3 \left[-\frac{24}{7} \dot{r}\, x_{\langle i} x_{j\rangle} +\frac{48}{7} r\,  x_{\langle i} \dot{x}_{j\rangle} \right]\,.
\end{equation}
All expressions needed to reach a 3PN accurate waveform are known for both the binary trajectory and the other SFT tensors $I_L,\ldots,Z_L$~\cite{Mishra:2015bqa,Cho:2018upo,Bini:2022enm}. Notice that, to leading PM order, the near zone multipoles include a $G$-independent contribution that is obtained by substituting the free motion in the trajectory. This means that contributions with an explicit factor of $G^2$, such as the one in~\eqref{I2ret}, are relevant for the comparison with the NLO PM waveform thus providing a non-trivial check with several non-linear effects that appear already in the near zone analysis.

The contributions coming from the instantaneous part (see~\eqref{I2loc} for the quadrupole) give integer PN corrections relative to the leading term, while~\eqref{I2ret} yields half-integer PN corrections that combine with those discussed in the subsections below. So for the time being we focus on the first type of contribution and would like to highlight an interesting pattern.
At the Newtonian level, which should be compared with~\eqref{eq:U2LO} and~\eqref{LONLOrelationPi}, we can use $K\approx 1$ and so the near zone multiples depend on $\tilde\phi$ in \eqref{eq:QKpb} only through trigonometric functions. Instead at the 1PN level, one needs to use the more accurate expression $K\approx 1 + 3 G^2 m^2/(b^2 p_\infty^2)$ and keep track of the contributions coming from expansions such as $\sin\phi\approx \sin\tilde\phi + (K-1) \tilde\phi \cos\tilde\phi$. Since the correction term is of order $G^2$, for our purposes one can then use the free trajectory in all the other objects appearing in this calculation. The Fourier transform of the terms proportional to $(K-1)$ can be recast in terms of (derivatives of)~\eqref{eq:FTBesselPN} with {\em integer} values of $\alpha$. Thus these contributions can be written in terms of elementary functions and so have a different structure from all the other 1PN corrections in~\eqref{I2loc}. Then it is possible to make a finer comparison between the MPM and the amplitudes results: the radiative multipoles obtained from~\eqref{eq:B1odd(i)} reproduce all the 1PN terms that involve Bessel function of integer order, while the contributions we have just discussed match those obtained from $\mathcal B_{1E}$ on the amplitudes side.

A similar pattern seems to hold at 2PN as well. It is non-trivial to check for instance the contribution to $I^{\rm inst}_{33}$ that can be written in terms of elementary functions and starts at 2PN order  $I^{\rm inst}_{33}\approx - G^2 m^3 \pi \nu (5-2\nu) b/p_\infty e^{-u}(1+u)/u+\ldots$ (where the dots stand for the terms with Bessel functions of integer order). On the MPM side, this result is obtained with a mechanism similar to the one discussed at 1PN level, the only difference being that now the origin of the polynomial dependence on $\tilde\phi$ is via the correction parameterized by $f_t$ in~\eqref{eq:QKpc}. Again one can use the free trajectory in evaluating this correction and the Fourier transform to the frequency domain is done by using~\eqref{eq:FTBesselPN} with {\em integer} values of $\alpha$. The MPM result matches~\eqref{eq:56d} derived in the amplitudes framework. 

\subsection{Nonlinear effects outside the near zone}
Other nonlinear contributions arise from the matching between the near zone and the far-zone multipoles following the approach schematically discussed at the beginning of this section. 
We collect here the formulas determining them \cite{Blanchet:1997ji,Blanchet:1996wx} and provide their explicit expressions in the ancillary file (see~\cite{Blanchet:2013haa,Mishra:2015bqa} for more details).

To the order of interest here, $U_{ijkl}$ only receives the nonlinear contribution
\begin{equation}\label{eq:U4QQ}
	U_{ijkl}^{QQ}
	=
	-\frac{G}{c^3}
	\left[
	\frac{21}{5}\,
	I_{\langle ij}^{\phantom{(5)}} I_{k\ell\rangle}^{(5)}
	+
	\frac{63}{5}\,
	I_{\langle ij}^{(1)} I_{k\ell\rangle}^{(4)}
	+
	\frac{102}{5}\, I_{\langle ij}^{(2)} I_{k\ell\rangle}^{(3)}
	\right].
\end{equation}
Here and in the following, $f^{(n)}(t)=d^nf(t)/dt^n$.
Note that, since we restrict to $\mathcal O (G^2)$ corrections to the multipoles, Eq.~\eqref{eq:U4QQ} only receives contributions from interference terms between the $G$-independent term in the source quadrupole $I_{ij}$, which involves terms up to $t^2$, and the dynamical $\mathcal O(G)$ quadrupole. A similar pattern holds for most of the contributions below as well.

Turing to $V_{ijk}$, it receives both a contribution due to nonlinearities
\begin{equation}\label{}
	V_{ijk}^{QQ}
	=
	\frac{G}{c^3}
	\left[
	\frac{1}{10}\,
	\epsilon_{ab\langle i }
	I_{j}^{(5)a} I_{k\rangle b}^{\phantom{a}}
	-
	\frac{1}{2}
	\epsilon_{ab\langle i}
	 I_{j}^{(4)a} I_{k\rangle b}^{(1)}
	\right]
\end{equation}
and one due to angular momentum,
\begin{equation}\label{}
	V_{ijk}^{LQ}
	=
	- \frac{2G}{c^3}\, I_{\langle ij}^{(4)} L^{\phantom{(4)}}_{k\rangle}
\end{equation}
with
\begin{equation}\label{}
	L_k = \delta_{kz} p b\,.
\end{equation}

The quadrupole receives a correction due to nonlinearities,
\begin{equation}\label{eq:U2QQ}
	U_{ij}^{QQ} 
	=
	\frac{G}{c^5}\left[
	\frac{1}{7}\,I_{a\langle i}^{(5)}I_{j\rangle a}
	-
	\frac{5}{7}\, I_{a\langle i}^{(4)} I_{j\rangle a}^{(1)}
	-
	\frac{2}{7}\, I_{a\langle i}^{(3)} I_{j\rangle a}^{(2)}
	\right],
\end{equation}
one due to the gauge multipole moments~\cite{Blanchet:1996wx},
\begin{equation}\label{eq:U2WQ}
	U_{ij}^{WQ} 
	=
	\frac{4G}{c^5}\left[
	W^{(2)} I_{ij} - W^{(1)}I_{ij}^{(2)}
	\right]^{(2)},
\end{equation}
with
\begin{equation}\label{}
	W = \frac{1}{3}m\nu\, r\,\dot{r}\,,
\end{equation}
and an angular-momentum dependent contribution,
\begin{equation}\label{eq:U2LQ}
	U_{ij}^{LQ} = \frac{1}{3}\,\frac{G}{c^5}\,\epsilon_{ab\langle i}^{\phantom{(4)}} I_{j\rangle a}^{(4)} L_b\,.
\end{equation}
Note that 
we need to retain its the first correction, $\mathcal O(G)$, to the trajectory when evaluating the square parenthesis of \eqref{eq:U2WQ}, since the leading term is time independent and thus it drops out from the final result.
Additional sources of $\mathcal O(\nu^2)$ fractional PN effects come from the already discussed retardation effect \eqref{I2loc} and from the radiation-reaction contribution in \eqref{eq:U2RR} below.

\subsection{Tail formula and radiation reaction effects}
Time antisymmetric contributions also arise from tail and radiation reaction effects. Tail terms capture the rescattering of the emitted radiation on the Coulombic field produced by the source, and take into account the fact that the near-zone light cones do not coincide with the flat ones in the wave zone. 
They are specific to the nonlinear nature of gravity, and for instance they would be absent in electromagnetism.
The tail formula that incorporates these effects reads
\begin{subequations}\label{eq:tmpm}
\begin{align}
\mathrm{U}_L^\text{tail} 
&= 
\frac{2GE}{c^3}  i\omega  \mathrm{U}_L^\text{tree} \left(\log({2b_0\omega})-\kappa_\ell + \gamma-\frac{i\pi}{2}\right)\!,
\label{eq:tmpma} \\
\mathrm{V}_L^\text{tail} 
&= 
\frac{2GE}{c^3}  i\omega  \mathrm{V}_L^\text{tree} \left(\log(2b_0\omega)-\pi_\ell + \gamma-\frac{i\pi}{2}\right)\!,
\label{eq:tmpmb}
\end{align}
\end{subequations}
where the shift involving the Euler--Mascheroni constant $\gamma$ and the $\frac{i\pi}{2}$ comes from the Fourier transform 
to frequency domain. We immediately see that the terms with $\frac{i\pi}{2}$ in \eqref{eq:tmpm} match $\mathcal B_{1O}^{(h)}$ in \eqref{eq:B1odd(h)} to all orders.

To facilitate the matching with amplitude-based results for the terms without the extra power of $\pi$, we find it convenient to trade the cutoff $b_0$ for $\mu_\text{IR}$ as follows, 
\begin{equation}\label{eq:redefb0}
	\log(2b_0\mu_\text{IR})= - \gamma\,,
\end{equation}
thereby absorbing the $\gamma$. Note that accounting for the exponentiation of infrared divergences according to \eqref{eq:expIRdiv} makes the matching of scheme choices \eqref{eq:redefb0} simpler compared to \cite{Georgoudis:2023eke}.
In \eqref{eq:tmpm}, the ``harmonic numbers'' $\kappa_\ell$ and $\pi_\ell$ are given  by
\begin{align}\label{}
	\kappa_\ell
	&= \frac{2\ell^2+5\ell+4}{\ell(\ell+1)(\ell+2)} + \sum_{k=1}^{\ell-2}\frac{1}{k}\,,
	\\
	\pi_\ell
	&= \frac{\ell-1}{\ell(\ell+1)} + \sum_{k=1}^{\ell-1}\frac{1}{k}\,.
\end{align}
In particular, since we consider multipole moments up to $U_4$, we are going to need
\begin{equation}\label{}
	\kappa_2 = \frac{11}{12}\,,\qquad
	\kappa_3 = \frac{97}{60}\,,\qquad
	\kappa_4 = \frac{59}{30}
\end{equation}
and
\begin{equation}\label{}
	\pi_2 = \frac{7}{6}\,,\qquad
	\pi_3 = \frac{5}{3}\,. 
\end{equation}
In the probe limit, i.e.~at leading order in the small $\nu$ limit, the tail contributions~\eqref{eq:tmpm} yield all the corrections of relative half-integer PN order for each multipole. This is a rather nontrivial statement on the amplitude side, where such corrections come from the (challenging) expansion of the Compton cuts.

Finally, we consider a contribution due to the radiation-reacted trajectory.
This is due to the leading-order radiation-reaction force (see e.g.~\cite{Bini:2021gat,Manohar:2022dea})
\begin{equation}\label{}
	f_{RR}^\mu = \frac{8G^2m^3p_\infty^3\nu^2}{5 r^5}\left(p_\infty t(3 b_e^\mu+2p_\infty t\,e^\mu)-b^2 e^\mu\right),
\end{equation}
where we can restrict to the straight-line motion, so that in particular $r = \sqrt{b^2+p_\infty^2 t^2}$,
and translates into a correction to the trajectories, 
\begin{equation}\label{eq:xRR}
	x_{RR}^\mu = \frac{8G^2m^2p_\infty \nu}{5 b^2 r}\left(
 	b^2 e^\mu - (r+p_\infty t) b^\mu_e
	\right).
\end{equation}
Note that $x_{RR}^\mu(t=-\infty)=0$, while
\begin{equation}\label{}
	x_{RR}^\mu(t=+\infty) = -\frac{16G^2 m^2p_\infty \nu}{5 b^2}\,b_e^\mu = (\Delta b) \,\frac{b_e^\mu}{b}\,,
\end{equation}
so that this correction shrinks the impact parameter, $\Delta b<0$. This leads to the $\mathcal O(G^2)$ loss of mechanical angular momentum $\Delta L = p\Delta b\simeq m p_\infty \nu \Delta b$, which matches the leading PN limit of the results obtained in \cite{Damour:2020tta,Manohar:2022dea,DiVecchia:2022owy}.
The sought-for correction to the quadrupole then reads as follows,
\begin{equation}\label{eq:U2RR}
	U_{ij}^{RR} = 2m 
	\nu\left(
	{x}_{\langle i}^{\phantom{R}}\, {x}_{j\rangle}^{RR}
	\right)^{(2)},
\end{equation}
where $x_i^{RR}$ is given by \eqref{eq:xRR} and $x_i$ can be approximated with the free trajectory, $x_i \simeq (b,p_\infty t,0)$.

An interesting check is to calculate on the MPM side the 2.5PN correction to the radiative quadrupole, where one needs to combine the contributions of~\eqref{I2ret}, \eqref{eq:U2QQ}, \eqref{eq:U2WQ}, \eqref{eq:U2LQ} and~\eqref{eq:U2RR} together with the (PN expansion of the) tail term~\eqref{eq:tmpma}. The result matches precisely~\eqref{eq:63} providing a non-trivial test of both the MPM and the amplitude-based calculations.

\section{Comparison between Amplitude and MPM Results}
\label{sec:comparison}

Let us now compare the results obtained by expanding the amplitude-based waveform in Section~\ref{sec:PNampl} with those calculated using the MPM-PN approach in Section~\ref{sec:MPM}. 

For $\mathcal A_0$, $\mathcal B_{1O}$ and $\mathcal B_{1E}$, for which Ref.~\cite{Bini:2023fiz} had already found agreement between the two methods, we obtained the complete multipole decomposition up to NNNLO PN order on the amplitude side, while we only calculated a few instructive multipole components on the MPM side as cross-checks. We find perfect agreement both for sample components involving Bessel functions $K_{0}(u)$, $K_1(u)$ and for those involving $e^{-u}$, such as $U_{33}^\text{NLO}$ in \eqref{eq:56d}.

The tail logarithm \eqref{eq:Creg} is of course exactly the same on both sides. Finally, for the quantity $C^{\text{reg}}$, we performed several checks, comparing all multipoles up to NNLO, i.e.~2.5PN order, between the two approaches.
While this quantity appears ``in one block'' as the sum of Compton cuts in the amplitude framework, in MPM it is given by the sum of tail terms as well as nonlinear, gauge and radiation-reaction contributions, which we summarized in the subsections of Section~\ref{sec:MPM} and are also collected in the ancillary file.
The sum of all of these contributions matches exactly the prediction obtained for $C^{\text{reg}}$ from the amplitude side, but this check requires paying special attention to two important points.

The first point is that the MPM results hold in the BMS supertranslation frame where the asymptotic shear has a nontrivial $\mathcal O(G)$ dictated by the free trajectories \cite{Damour:2020tta}. On the other hand, $\omega>0$ amplitudes do not give rise to this term, and lead to a different BMS frame. One thus needs to move from the latter to the former, via \eqref{eq:ST} \cite{Veneziano:2022zwh,Georgoudis:2023eke}.  
The second point is that the subtraction of infrared divergences also induces a modification of the finite term via \eqref{eq:c1regDEF}, where nontrivial $\epsilon/\epsilon$ terms appear by consistency with dimensional regularization.

Interestingly, all one-loop integer PN corrections involving $K_0(u)$, $K_1(u)$ are actually fixed in terms of tree-level multipoles via \eqref{eq:B1odd(i)}. This simple relation, which originates from the full PM amplitude calculation, is rather surprising from the MPM-PN perspective. Vice-versa, the tail formula that emerges naturally in the MPM framework guarantees that the $\mathcal O(\nu)$ part of the time-reversal odd corrections to each one-loop multipole is in fact proportional to the corresponding tree-level one, the relative factor being simply fixed by the corresponding ``harmonic number''. Naively, one would not expect such a dramatic simplification by simply looking at the Compton cuts of the amplitude.
Finally, the agreement obtained here for the 2.5PN correction to the quadrupole is an extremely nontrivial test of both formalisms, as it involves a rather technically complicated, albeit conceptually straightforward, expansion of $C^\text{reg}$ on one side, and the calculation of several physically distinct contributions on the other \cite{Mishra:2015bqa}. 

It will be interesting to test the agreement between these nicely complementary methods also beyond the case of scalar point-like objects, including also tidal effects and spin corrections \cite{Bohnenblust:2023qmy}, to see whether the supertranslation frame plays an analogous role, as we expect. 
Another important direction consists of course in increasing the precision, both in the velocity expansion and in the PM order.
Obtaining high orders in $\lambda$ of the tree level kernel is not difficult, of course, and several additional multipoles could be obtained by more efficiently automatizing their extraction and the Fourier transforms.
Moreover, we expect the method developed here, based on fitting a polynomial ansatz of the PN limit achieved by a numerical sampling of the expansion, to be easily applicable to obtain a few additional orders in the $\lambda$ expansion at one loop. Further mileage could be gained by implementing more advanced rational reconstruction methods (see e.g.~\cite{Peraro:2016wsq}).
The next order in $G$, instead, represents a nontrivial future challenge, as it requires
the classical $2\to3$ amplitude at two loops.
On the other side, the latest achievement of the MPM-PN formalism consists in the derivation of the gravitational-wave phase and frequency to 4.5PN order beyond the leading quadrupole formula \cite{Blanchet:2023bwj,Blanchet:2023sbv}.
Finally it would of course be very interesting to study the connection between the scattering and the bound waveforms at the subleading PM order, building on the recent results of~\cite{Adamo:2024oxy}. 

\subsection*{Acknowledgments}
We are grateful to Donato Bini, William J. Torres Bobadilla, Thibault Damour, Stefano De Angelis, Andrea Geralico, Donal O'Connell and Fei Teng for helpful discussions.
A.~G. is supported by a Royal Society funding, URF\textbackslash{R}\textbackslash221015. C.~H. is supported by UK Research and Innovation (UKRI) under the UK government’s Horizon Europe funding guarantee [grant EP/X037312/1 ``EikoGrav: Eikonal Exponentiation and Gravitational Waves'']. R.~R. is partially supported by the UK EPSRC grant ``CFT and Gravity: Heavy States and Black Holes'' EP/W019663/1 and the STFC grants ``Amplitudes, Strings and Duality'', grant numbers ST/T000686/1 and ST/X00063X/1. R. R. would like to thank IHES for hospitality during the initial stages of this project. No new data were generated or analysed during this study.

\appendix

\section{Multipole Expansion}
\label{sec:multipoles}

Here, we summarize the multipole expansion of the waveform which is commonly employed in the PN literature (see e.g.~\cite{Blanchet:1985sp,Blanchet:2013haa}).
To this end, let us first define the traceless projection of its spatial part in the center-of-mass frame,
\begin{equation}\label{eq:WaveformPMtraceless}
	{\mathcal{H}}_{ij}(\omega n)
	=
	\frac{1}{\kappa}\left[
	\tilde{W}^{\text{reg}}_{ij}(\omega n)-\frac13 \delta_{ij} \tilde{W}^\text{reg}_{kk}(\omega n)\right]
\end{equation}
and further consider its transverse-traceless (TT) projection, 
\begin{equation}\label{eq:TTprojH}
	\mathcal H_{ij}^\text{TT} 
	= 
	\mathcal P_{ijab}(n)
	\mathcal H_{ij}
\end{equation}
where $\mathcal P_{ijab}(n)$ is the TT projector defined with respect to $n_i$, that is,
\begin{equation}\label{}
	\begin{split}
		\mathcal P_{ij}(n) &= \delta_{ij}-n_i n_j\,,\\
		\mathcal P_{ijab}(n) &= 
		\mathcal P_{i(a}(n)\mathcal P_{b)j}(n)
		-
		\frac{1}{2}
		\mathcal P_{ij}(n)\mathcal P_{ab}(n)\,.
	\end{split}
\end{equation}
Owing to the TT projection in \eqref{eq:TTprojH}, $\mathcal H_{ij}^\text{TT}$ descends to a symmetric traceless tensor on the two-sphere described by $n^i$. Letting $x^A$ with $A=1,2$ denote coordinates thereon, and $e_A^i = \partial_A n^i$ a basis of tangent vectors, this tensor is
\begin{equation}\label{}
	\mathcal H_{AB} 
	=  e_A^i e_B^j \mathcal H_{ij} + \frac{1}{2}\,\gamma_{AB}\, n^i n^j \mathcal H_{ij}\,,
\end{equation}
where $\gamma_{AB} = e_A \cdot e_B$ is the metric on the two-sphere.

One can then decompose $\mathcal H_{AB}$ in terms of a basis of symmetric traceless tensors with definite eigenvalue under the action of the Laplacian on the sphere, $\Delta = {D}_A {D}^A$, with $D_A$ the covariant derivative associated to $\gamma_{AB}$.
These can be obtained by taking derivatives of irreducible scalar $\mathcal{K}^{(\ell)}$ and vector harmonics $\mathcal K^{(\ell)}_A$, with $D^A \mathcal K_A^{(\ell)}=0$, which are such that
\begin{subequations}
 \begin{align}
 	\label{eq:Kell}
 	\Delta\mathcal{K}^{(\ell)}
 	&=
 	-\ell(\ell+1)\,\mathcal{K}^{(\ell)}\,,\\
 	\label{eq:KAell}
 	\Delta\mathcal{K}^{(\ell)}_A
 	&=
 	-[\ell(\ell+1)-1]\,\mathcal{K}_A^{(\ell)}\,.
 \end{align}
\end{subequations}
Indeed, $\mathrm{U}_{AB}^{(\ell)} = (2D_AD_B-\gamma_{AB}\Delta) \mathcal K^{(\ell)}$ and $\mathrm{V}_{AB}^{(\ell)}=D_{(A}K_{B)}^{(\ell)}$ obey
\begin{subequations}
	\begin{align}
		\label{eq:UAB}
		\Delta \mathrm{U}^{(\ell)}_{AB}
		&=
		-[\ell(\ell+1)-4]\,\mathrm{U}^{(\ell)}_{AB}\,,\\
		\label{eq:VAB}
		\Delta \mathrm{V}^{(\ell)}_{AB}
		&=
		-[\ell(\ell+1)-4]\,\mathrm{V}^{(\ell)}_{AB}\,.
	\end{align}
\end{subequations}
These provide an \textit{orthogonal} basis of symmetric traceless rank-two on the two-sphere, e.g. $\oint U_{AB}^{(\ell)}U^{(\ell')AB}d\Omega=0$ if $\ell\neq\ell'$ and $\oint U_{AB}^{(\ell)}V^{(\ell')AB}d\Omega=0$ for any $\ell$, $\ell'$, where $d\Omega= \sqrt{\gamma}\, d^2x$ and indices are raised using $\gamma^{AB}$.

Explicit representations for $\mathrm{U}^{(\ell)}_{AB}$ and $\mathrm{V}^{(\ell)}_{AB}$ thus follow from those of $\mathcal{K}^{(\ell)}$, $\mathcal{K}_A^{(\ell)}$, which can be constructed as follows,
\begin{equation}\label{eq:explKKA}
		\mathcal{K}^{(\ell)}
		=
		C_{L} n_L
		\,,\qquad
		\mathcal{K}_A^{(\ell)}
		=
		e_A^a \epsilon_{abc} C_{b L-1} n_{c L-1}\,,
\end{equation}
where $C_L$ are ($n$-independent) symmetric trace-free (STF) tensors
and 
a capital letter $L$ stands for a multi-index with $\ell$ entries, for instance
\begin{equation}\label{}
	C_{L}=C_{i_1i_2\cdots i_\ell}\,,\qquad
	n_{cL-1} = n_{c}n_{i_1}\cdots n_{i_{\ell-1}}\,.
\end{equation}
A convenient orthogonal basis for the tensors $C_{L}$ is given by the STF projection of the $2\ell+1$ monomials $(e_x\pm ie_y)^{i_1}\cdots (e_x\pm ie_y)^{i_m}e_{z}^{i_{m+1}}\cdots e_z^{i_\ell}$ for $m\le \ell$.
When taking derivatives of \eqref{eq:explKKA} to obtain $\mathrm{U}_{AB}^{(\ell)}$ and $\mathrm{V}^{(\ell)}_{AB}$, it is convenient to recall the property 
\begin{equation}\label{eq:DADBgammaAB}
D_A D_B n^i = - \gamma_{AB} n^i\,.
\end{equation} 
Let us also remark that, while $\mathcal{K}^{(\ell)}$, $\mathcal K_A^{(\ell)}$ satisfying \eqref{eq:Kell}, \eqref{eq:KAell} exist for $\ell\ge0$ and $\ell\ge1$, respectively, the resulting $\mathrm{U}_{AB}^{(\ell)}$ and $\mathrm{V}^{(\ell)}_{AB}$ satisfying \eqref{eq:UAB}, \eqref{eq:VAB} are only nonzero for $\ell\ge2$.

Using the identity \eqref{eq:DADBgammaAB}, in view of the above discussion, we thus see that the spectral waveform
admits the following unique decomposition in terms of STF tensors  $\mathrm{U}_{L}(u)$, $\mathrm{V}_{L}(u)$ \cite[Eq.~(66)]{Blanchet:2013haa}
\begin{widetext}
\begin{equation}\label{Blanchet66}
	\begin{split}
		\mathcal H_{ij}^\text{TT} = 
		\mathcal P_{ijab}(n)\sum_{\ell=2}^\infty\frac{1}{\ell!}
		\Bigg[
		n_{L-2}\, \mathrm{U}_{abL-2}(u)
		-\frac{2\ell}{\ell+1}\,n_{cL-2}\epsilon_{cd(a} \, \mathrm{V}_{b)dL-2}(u)
		\Bigg]
	\end{split}
\end{equation}
\end{widetext}
Note that we include the symmetry factor in the symmetrization of spatial (lowercase Latin) indices
$A_{(i}B_{j)} = \frac12 A_i B_j + \frac 12 A_j B_i$.
Importantly, the multipole moments $\mathrm{U}_L(u)$ and $\mathrm{V}_L(u)$ only depend on the (dimensionless) frequency $u$ in \eqref{eq:uKT} and not on the angles.
A key simplification is that, order by order in the PN expansion, only the first few terms in the sum appearing in Eq.~\eqref{Blanchet66} actually show up. In practice, once a basis is written down explicitly following the above steps, one can simply contract \eqref{Blanchet66} with the desired harmonic and perform the integral to extract it from the expression.

In the text, we perform the decomposition \eqref{Blanchet66} both for the tree-level amplitude  and for the building blocks of the one-loop ``eikonal'' waveform kernel \eqref{eq:Weik}. For the latter we employ the notation $\mathrm U_{X L}$, $\mathrm V_{X L}$, where $X\mapsto O$ for $\mathcal B_{1O}$ in \eqref{eq:B1odd} and $X\mapsto E$ for $\mathcal B_{1E}$ in \eqref{eq:B1even}, $X\mapsto\text{IR}$ for the coefficient of the IR-running logarithm in \eqref{eq:Creg} and $X\mapsto C$ for $C^\text{reg}$ also in \eqref{eq:Creg}. In the ancillary files, we further distinguish between $\mathcal B_{1O}^{(i)}$ and $\mathcal B_{1O}^{(h)}$ in \eqref{eq:B1odd(i)}, \eqref{eq:B1odd(h)}.

We conclude this appendix by recalling the formula for the energy flux in terms of the MPM expansion~\cite{Blanchet:2013haa}, which in the frequency-domain is
\begin{widetext}
\begin{equation}
    \label{eq:dEdo}
      \frac{dE_\text{rad}}{d\omega} = \frac{G}{\pi} \sum_{\ell=2}^{+\infty} \left\{\frac{(\ell+1)(\ell+2)\,\omega^2}{(\ell-1) \ell \ell !(2 \ell+1) ! !} \mathrm{U}^*_L \mathrm{U}_L  + \frac{4 \ell(\ell+2)\,\omega^2}{(\ell-1)(\ell+1) !(2 \ell+1) ! !} \mathrm{V}^*_L \mathrm{V}_L\right\},
    \end{equation}
and the formula for the flux of the radiated linear momentum
  \begin{equation}
  \label{eq:dPdo}
\frac{d P_{i\,{\mathrm{rad}}}}{d\omega} = \frac{G}{\pi} \sum_{\ell=2}^{\infty}\operatorname{Re}\left[ \frac{2(\ell+2)(\ell+3)\, \omega^2}{\ell(\ell+1) !(2 \ell+3) ! !} \mathrm{U}_{i L}^* \mathrm{U}_L +\frac{8(\ell+3)\, \omega^2}{(\ell+1) !(2 \ell+3) ! !} \mathrm{V}^*_{i L} \mathrm{V}_L + \frac{8(\ell+2) \, \omega^2 \,\epsilon_{i a b} \mathrm{U}^*_{a L-1} \mathrm{V}_{b L-1}}{(\ell-1)(\ell+1) !(2 \ell+1) ! !}  \right].
\end{equation}
\end{widetext}

\providecommand{\href}[2]{#2}\begingroup\raggedright\endgroup


\begin{thebibliography}{10}
	
	\bibitem{Georgoudis:2023eke}
	A.~Georgoudis, C.~Heissenberg, and R.~Russo, ``{An eikonal-inspired approach to
		the gravitational scattering waveform},''
	\href{http://dx.doi.org/10.1007/JHEP03(2024)089}{{\em JHEP} {\bf 03} (2024)
		089}, \href{http://arxiv.org/abs/2312.07452}{{\tt arXiv:2312.07452
			[hep-th]}}.
	
	\bibitem{Bern:2019nnu}
	Z.~Bern, C.~Cheung, R.~Roiban, C.-H. Shen, M.~P. Solon, and M.~Zeng,
	``{Scattering Amplitudes and the Conservative Hamiltonian for Binary Systems
		at Third Post-Minkowskian Order},''
	\href{http://dx.doi.org/10.1103/PhysRevLett.122.201603}{{\em Phys. Rev.
			Lett.} {\bf 122} (2019) no.~20, 201603},
	\href{http://arxiv.org/abs/1901.04424}{{\tt arXiv:1901.04424 [hep-th]}}.
	
	\bibitem{Bern:2019crd}
	Z.~Bern, C.~Cheung, R.~Roiban, C.-H. Shen, M.~P. Solon, and M.~Zeng, ``{Black
		Hole Binary Dynamics from the Double Copy and Effective Theory},''
	\href{http://dx.doi.org/10.1007/JHEP10(2019)206}{{\em JHEP} {\bf 10} (2019)
		206}, \href{http://arxiv.org/abs/1908.01493}{{\tt arXiv:1908.01493
			[hep-th]}}.
	
	\bibitem{Bern:2021yeh}
	Z.~Bern, J.~Parra-Martinez, R.~Roiban, M.~S. Ruf, C.-H. Shen, M.~P. Solon, and
	M.~Zeng, ``{Scattering Amplitudes, the Tail Effect, and Conservative Binary
		Dynamics at O(G4)},''
	\href{http://dx.doi.org/10.1103/PhysRevLett.128.161103}{{\em Phys. Rev.
			Lett.} {\bf 128} (2022) no.~16, 161103},
	\href{http://arxiv.org/abs/2112.10750}{{\tt arXiv:2112.10750 [hep-th]}}.
	
	\bibitem{Bern:2021dqo}
	Z.~Bern, J.~Parra-Martinez, R.~Roiban, M.~S. Ruf, C.-H. Shen, M.~P. Solon, and
	M.~Zeng, ``{Scattering Amplitudes and Conservative Binary Dynamics at ${\cal
			O}(G^4)$},'' \href{http://dx.doi.org/10.1103/PhysRevLett.126.171601}{{\em
			Phys. Rev. Lett.} {\bf 126} (2021) no.~17, 171601},
	\href{http://arxiv.org/abs/2101.07254}{{\tt arXiv:2101.07254 [hep-th]}}.
	
	\bibitem{Cheung:2018wkq}
	C.~Cheung, I.~Z. Rothstein, and M.~P. Solon, ``{From Scattering Amplitudes to
		Classical Potentials in the Post-Minkowskian Expansion},''
	\href{http://dx.doi.org/10.1103/PhysRevLett.121.251101}{{\em Phys. Rev.
			Lett.} {\bf 121} (2018) no.~25, 251101},
	\href{http://arxiv.org/abs/1808.02489}{{\tt arXiv:1808.02489 [hep-th]}}.
	
	\bibitem{KoemansCollado:2019ggb}
	A.~Koemans~Collado, P.~Di~Vecchia, and R.~Russo, ``{Revisiting the second
		post-Minkowskian eikonal and the dynamics of binary black holes},''
	\href{http://dx.doi.org/10.1103/PhysRevD.100.066028}{{\em Phys. Rev. D} {\bf
			100} (2019) no.~6, 066028}, \href{http://arxiv.org/abs/1904.02667}{{\tt
			arXiv:1904.02667 [hep-th]}}.
	
	\bibitem{Cheung:2020gyp}
	C.~Cheung and M.~P. Solon, ``{Classical gravitational scattering at $
		\mathcal{O} $(G$^{3}$) from Feynman diagrams},''
	\href{http://dx.doi.org/10.1007/JHEP06(2020)144}{{\em JHEP} {\bf 06} (2020)
		144}, \href{http://arxiv.org/abs/2003.08351}{{\tt arXiv:2003.08351
			[hep-th]}}.
	
	\bibitem{Cristofoli:2020uzm}
	A.~Cristofoli, P.~H. Damgaard, P.~Di~Vecchia, and C.~Heissenberg,
	``{Second-order Post-Minkowskian scattering in arbitrary dimensions},''
	\href{http://dx.doi.org/10.1007/JHEP07(2020)122}{{\em JHEP} {\bf 07} (2020)
		122}, \href{http://arxiv.org/abs/2003.10274}{{\tt arXiv:2003.10274
			[hep-th]}}.
	
	\bibitem{DiVecchia:2020ymx}
	P.~Di~Vecchia, C.~Heissenberg, R.~Russo, and G.~Veneziano, ``{Universality of
		ultra-relativistic gravitational scattering},''
	\href{http://dx.doi.org/10.1016/j.physletb.2020.135924}{{\em Phys. Lett. B}
		{\bf 811} (2020)  135924}, \href{http://arxiv.org/abs/2008.12743}{{\tt
			arXiv:2008.12743 [hep-th]}}.
	
	\bibitem{Damour:2020tta}
	T.~Damour, ``{Radiative contribution to classical gravitational scattering at
		the third order in $G$},''
	\href{http://dx.doi.org/10.1103/PhysRevD.102.124008}{{\em Phys. Rev. D} {\bf
			102} (2020) no.~12, 124008}, \href{http://arxiv.org/abs/2010.01641}{{\tt
			arXiv:2010.01641 [gr-qc]}}.
	
	\bibitem{DiVecchia:2021ndb}
	P.~Di~Vecchia, C.~Heissenberg, R.~Russo, and G.~Veneziano, ``{Radiation
		Reaction from Soft Theorems},''
	\href{http://dx.doi.org/10.1016/j.physletb.2021.136379}{{\em Phys. Lett. B}
		{\bf 818} (2021)  136379}, \href{http://arxiv.org/abs/2101.05772}{{\tt
			arXiv:2101.05772 [hep-th]}}.
	
	\bibitem{DiVecchia:2021bdo}
	P.~Di~Vecchia, C.~Heissenberg, R.~Russo, and G.~Veneziano, ``{The eikonal
		approach to gravitational scattering and radiation at $ \mathcal{O}
		$(G$^{3}$)},'' \href{http://dx.doi.org/10.1007/JHEP07(2021)169}{{\em JHEP}
		{\bf 07} (2021)  169}, \href{http://arxiv.org/abs/2104.03256}{{\tt
			arXiv:2104.03256 [hep-th]}}.
	
	\bibitem{Herrmann:2021tct}
	E.~Herrmann, J.~Parra-Martinez, M.~S. Ruf, and M.~Zeng, ``{Radiative classical
		gravitational observables at $ \mathcal{O} $(G$^{3}$) from scattering
		amplitudes},'' \href{http://dx.doi.org/10.1007/JHEP10(2021)148}{{\em JHEP}
		{\bf 10} (2021)  148}, \href{http://arxiv.org/abs/2104.03957}{{\tt
			arXiv:2104.03957 [hep-th]}}.
	
	\bibitem{Bjerrum-Bohr:2021vuf}
	N.~E.~J. Bjerrum-Bohr, P.~H. Damgaard, L.~Plant\'e, and P.~Vanhove,
	``{Classical gravity from loop amplitudes},''
	\href{http://dx.doi.org/10.1103/PhysRevD.104.026009}{{\em Phys. Rev. D} {\bf
			104} (2021) no.~2, 026009}, \href{http://arxiv.org/abs/2104.04510}{{\tt
			arXiv:2104.04510 [hep-th]}}.
	
	\bibitem{Bjerrum-Bohr:2021din}
	N.~E.~J. Bjerrum-Bohr, P.~H. Damgaard, L.~Plant\'e, and P.~Vanhove, ``{The
		amplitude for classical gravitational scattering at third Post-Minkowskian
		order},'' \href{http://dx.doi.org/10.1007/JHEP08(2021)172}{{\em JHEP} {\bf
			08} (2021)  172}, \href{http://arxiv.org/abs/2105.05218}{{\tt
			arXiv:2105.05218 [hep-th]}}.
	
	\bibitem{Damgaard:2021ipf}
	P.~H. Damgaard, L.~Plante, and P.~Vanhove, ``{On an exponential representation
		of the gravitational S-matrix},''
	\href{http://dx.doi.org/10.1007/JHEP11(2021)213}{{\em JHEP} {\bf 11} (2021)
		213}, \href{http://arxiv.org/abs/2107.12891}{{\tt arXiv:2107.12891
			[hep-th]}}.
	
	\bibitem{Brandhuber:2021eyq}
	A.~Brandhuber, G.~Chen, G.~Travaglini, and C.~Wen, ``{Classical gravitational
		scattering from a gauge-invariant double copy},''
	\href{http://dx.doi.org/10.1007/JHEP10(2021)118}{{\em JHEP} {\bf 10} (2021)
		118}, \href{http://arxiv.org/abs/2108.04216}{{\tt arXiv:2108.04216
			[hep-th]}}.
	
	\bibitem{Damgaard:2023vnx}
	P.~H. Damgaard, E.~R. Hansen, L.~Plant\'e, and P.~Vanhove, ``{The relation
		between KMOC and worldline formalisms for classical gravity},''
	\href{http://dx.doi.org/10.1007/JHEP09(2023)059}{{\em JHEP} {\bf 09} (2023)
		059}, \href{http://arxiv.org/abs/2306.11454}{{\tt arXiv:2306.11454
			[hep-th]}}.
	
	\bibitem{Damgaard:2023ttc}
	P.~H. Damgaard, E.~R. Hansen, L.~Plant\'e, and P.~Vanhove, ``{Classical
		observables from the exponential representation of the gravitational
		S-matrix},'' \href{http://dx.doi.org/10.1007/JHEP09(2023)183}{{\em JHEP} {\bf
			09} (2023)  183}, \href{http://arxiv.org/abs/2307.04746}{{\tt
			arXiv:2307.04746 [hep-th]}}.
	
	\bibitem{Kalin:2020mvi}
	G.~K\"alin and R.~A. Porto, ``{Post-Minkowskian Effective Field Theory for
		Conservative Binary Dynamics},''
	\href{http://dx.doi.org/10.1007/JHEP11(2020)106}{{\em JHEP} {\bf 11} (2020)
		106}, \href{http://arxiv.org/abs/2006.01184}{{\tt arXiv:2006.01184
			[hep-th]}}.
	
	\bibitem{Kalin:2020fhe}
	G.~K\"alin, Z.~Liu, and R.~A. Porto, ``{Conservative Dynamics of Binary Systems
		to Third Post-Minkowskian Order from the Effective Field Theory Approach},''
	\href{http://dx.doi.org/10.1103/PhysRevLett.125.261103}{{\em Phys. Rev.
			Lett.} {\bf 125} (2020) no.~26, 261103},
	\href{http://arxiv.org/abs/2007.04977}{{\tt arXiv:2007.04977 [hep-th]}}.
	
	\bibitem{Mogull:2020sak}
	G.~Mogull, J.~Plefka, and J.~Steinhoff, ``{Classical black hole scattering from
		a worldline quantum field theory},''
	\href{http://dx.doi.org/10.1007/JHEP02(2021)048}{{\em JHEP} {\bf 02} (2021)
		048}, \href{http://arxiv.org/abs/2010.02865}{{\tt arXiv:2010.02865
			[hep-th]}}.
	
	\bibitem{Dlapa:2021npj}
	C.~Dlapa, G.~K\"alin, Z.~Liu, and R.~A. Porto, ``{Dynamics of binary systems to
		fourth Post-Minkowskian order from the effective field theory approach},''
	\href{http://dx.doi.org/10.1016/j.physletb.2022.137203}{{\em Phys. Lett. B}
		{\bf 831} (2022)  137203}, \href{http://arxiv.org/abs/2106.08276}{{\tt
			arXiv:2106.08276 [hep-th]}}.
	
	\bibitem{Jakobsen:2022psy}
	G.~U. Jakobsen, G.~Mogull, J.~Plefka, and B.~Sauer, ``{All things retarded:
		radiation-reaction in worldline quantum field theory},''
	\href{http://dx.doi.org/10.1007/JHEP10(2022)128}{{\em JHEP} {\bf 10} (2022)
		128}, \href{http://arxiv.org/abs/2207.00569}{{\tt arXiv:2207.00569
			[hep-th]}}.
	
	\bibitem{Kalin:2022hph}
	G.~K\"alin, J.~Neef, and R.~A. Porto, ``{Radiation-reaction in the Effective
		Field Theory approach to Post-Minkowskian dynamics},''
	\href{http://dx.doi.org/10.1007/JHEP01(2023)140}{{\em JHEP} {\bf 01} (2023)
		140}, \href{http://arxiv.org/abs/2207.00580}{{\tt arXiv:2207.00580
			[hep-th]}}.
	
	\bibitem{Dlapa:2022lmu}
	C.~Dlapa, G.~K\"alin, Z.~Liu, J.~Neef, and R.~A. Porto, ``{Radiation Reaction
		and Gravitational Waves at Fourth Post-Minkowskian Order},''
	\href{http://dx.doi.org/10.1103/PhysRevLett.130.101401}{{\em Phys. Rev.
			Lett.} {\bf 130} (2023) no.~10, 101401},
	\href{http://arxiv.org/abs/2210.05541}{{\tt arXiv:2210.05541 [hep-th]}}.
	
	\bibitem{Dlapa:2023hsl}
	C.~Dlapa, G.~K\"alin, Z.~Liu, and R.~A. Porto, ``{Bootstrapping the
		relativistic two-body problem},''
	\href{http://dx.doi.org/10.1007/JHEP08(2023)109}{{\em JHEP} {\bf 08} (2023)
		109}, \href{http://arxiv.org/abs/2304.01275}{{\tt arXiv:2304.01275
			[hep-th]}}.
	
	\bibitem{Herrmann:2021lqe}
	E.~Herrmann, J.~Parra-Martinez, M.~S. Ruf, and M.~Zeng, ``{Gravitational
		Bremsstrahlung from Reverse Unitarity},''
	\href{http://dx.doi.org/10.1103/PhysRevLett.126.201602}{{\em Phys. Rev.
			Lett.} {\bf 126} (2021) no.~20, 201602},
	\href{http://arxiv.org/abs/2101.07255}{{\tt arXiv:2101.07255 [hep-th]}}.
	
	\bibitem{Manohar:2022dea}
	A.~V. Manohar, A.~K. Ridgway, and C.-H. Shen, ``{Radiated Angular Momentum and
		Dissipative Effects in Classical Scattering},''
	\href{http://dx.doi.org/10.1103/PhysRevLett.129.121601}{{\em Phys. Rev.
			Lett.} {\bf 129} (2022) no.~12, 121601},
	\href{http://arxiv.org/abs/2203.04283}{{\tt arXiv:2203.04283 [hep-th]}}.
	
	\bibitem{DiVecchia:2022owy}
	P.~Di~Vecchia, C.~Heissenberg, and R.~Russo, ``{Angular momentum of
		zero-frequency gravitons},''
	\href{http://dx.doi.org/10.1007/JHEP08(2022)172}{{\em JHEP} {\bf 08} (2022)
		172}, \href{http://arxiv.org/abs/2203.11915}{{\tt arXiv:2203.11915
			[hep-th]}}.
	
	\bibitem{Mougiakakos:2022sic}
	S.~Mougiakakos, M.~M. Riva, and F.~Vernizzi, ``{Gravitational Bremsstrahlung
		with Tidal Effects in the Post-Minkowskian Expansion},''
	\href{http://dx.doi.org/10.1103/PhysRevLett.129.121101}{{\em Phys. Rev.
			Lett.} {\bf 129} (2022) no.~12, 121101},
	\href{http://arxiv.org/abs/2204.06556}{{\tt arXiv:2204.06556 [hep-th]}}.
	
	\bibitem{Riva:2022fru}
	M.~M. Riva, F.~Vernizzi, and L.~K. Wong, ``{Gravitational bremsstrahlung from
		spinning binaries in the post-Minkowskian expansion},''
	\href{http://dx.doi.org/10.1103/PhysRevD.106.044013}{{\em Phys. Rev. D} {\bf
			106} (2022) no.~4, 044013}, \href{http://arxiv.org/abs/2205.15295}{{\tt
			arXiv:2205.15295 [hep-th]}}.
	
	\bibitem{DiVecchia:2022nna}
	P.~Di~Vecchia, C.~Heissenberg, R.~Russo, and G.~Veneziano, ``{The eikonal
		operator at arbitrary velocities I: the soft-radiation limit},''
	\href{http://dx.doi.org/10.1007/JHEP07(2022)039}{{\em JHEP} {\bf 07} (2022)
		039}, \href{http://arxiv.org/abs/2204.02378}{{\tt arXiv:2204.02378
			[hep-th]}}.
	
	\bibitem{Heissenberg:2022tsn}
	C.~Heissenberg, ``{Angular Momentum Loss due to Tidal Effects in the
		Post-Minkowskian Expansion},''
	\href{http://dx.doi.org/10.1103/PhysRevLett.131.011603}{{\em Phys. Rev.
			Lett.} {\bf 131} (2023) no.~1, 011603},
	\href{http://arxiv.org/abs/2210.15689}{{\tt arXiv:2210.15689 [hep-th]}}.
	
	\bibitem{Heissenberg:2023uvo}
	C.~Heissenberg, ``{Angular momentum loss due to spin-orbit effects in the
		post-Minkowskian expansion},''
	\href{http://dx.doi.org/10.1103/PhysRevD.108.106003}{{\em Phys. Rev. D} {\bf
			108} (2023) no.~10, 106003}, \href{http://arxiv.org/abs/2308.11470}{{\tt
			arXiv:2308.11470 [hep-th]}}.
	
	\bibitem{Kalin:2019rwq}
	G.~K{\"a}lin and R.~A. Porto, ``{From Boundary Data to Bound States},''
	\href{http://dx.doi.org/10.1007/JHEP01(2020)072}{{\em JHEP} {\bf 01} (2020)
		072}, \href{http://arxiv.org/abs/1910.03008}{{\tt arXiv:1910.03008
			[hep-th]}}.
	
	\bibitem{Kalin:2019inp}
	G.~K{\"a}lin and R.~A. Porto, ``{From boundary data to bound states. Part II.
		Scattering angle to dynamical invariants (with twist)},''
	\href{http://dx.doi.org/10.1007/JHEP02(2020)120}{{\em JHEP} {\bf 02} (2020)
		120}, \href{http://arxiv.org/abs/1911.09130}{{\tt arXiv:1911.09130
			[hep-th]}}.
	
	\bibitem{Saketh:2021sri}
	M.~V.~S. Saketh, J.~Vines, J.~Steinhoff, and A.~Buonanno, ``{Conservative and
		radiative dynamics in classical relativistic scattering and bound systems},''
	\href{http://dx.doi.org/10.1103/PhysRevResearch.4.013127}{{\em Phys. Rev.
			Res.} {\bf 4} (2022) no.~1, 013127},
	\href{http://arxiv.org/abs/2109.05994}{{\tt arXiv:2109.05994 [gr-qc]}}.
	
	\bibitem{Cho:2021arx}
	G.~Cho, G.~K\"alin, and R.~A. Porto, ``{From boundary data to bound states.
		Part III. Radiative effects},''
	\href{http://dx.doi.org/10.1007/JHEP04(2022)154}{{\em JHEP} {\bf 04} (2022)
		154}, \href{http://arxiv.org/abs/2112.03976}{{\tt arXiv:2112.03976
			[hep-th]}}. [Erratum: JHEP 07, 002 (2022)].
	
	\bibitem{Purrer:2019jcp}
	M.~P\"urrer and C.-J. Haster, ``{Gravitational waveform accuracy requirements
		for future ground-based detectors},''
	\href{http://dx.doi.org/10.1103/PhysRevResearch.2.023151}{{\em Phys. Rev.
			Res.} {\bf 2} (2020) no.~2, 023151},
	\href{http://arxiv.org/abs/1912.10055}{{\tt arXiv:1912.10055 [gr-qc]}}.
	
	\bibitem{Maggiore:2019uih}
	M.~Maggiore {\em et al.}, ``{Science Case for the Einstein Telescope},''
	\href{http://dx.doi.org/10.1088/1475-7516/2020/03/050}{{\em JCAP} {\bf 03}
		(2020)  050}, \href{http://arxiv.org/abs/1912.02622}{{\tt arXiv:1912.02622
			[astro-ph.CO]}}.
	
	\bibitem{Barausse:2020rsu}
	E.~Barausse {\em et al.}, ``{Prospects for Fundamental Physics with LISA},''
	\href{http://dx.doi.org/10.1007/s10714-020-02691-1}{{\em Gen. Rel. Grav.}
		{\bf 52} (2020) no.~8, 81}, \href{http://arxiv.org/abs/2001.09793}{{\tt
			arXiv:2001.09793 [gr-qc]}}.
	
	\bibitem{Ballmer:2022uxx}
	S.~W. Ballmer {\em et al.}, ``{Snowmass2021 Cosmic Frontier White Paper: Future
		Gravitational-Wave Detector Facilities},'' in {\em {Snowmass 2021}}.
	\newblock 3, 2022.
	\newblock \href{http://arxiv.org/abs/2203.08228}{{\tt arXiv:2203.08228
			[gr-qc]}}.
	
	\bibitem{Kovacs:1977uw}
	S.~J. Kovacs and K.~S. Thorne, ``{The Generation of Gravitational Waves. 3.
		Derivation of Bremsstrahlung Formulas},''
	\href{http://dx.doi.org/10.1086/155576}{{\em Astrophys. J.} {\bf 217} (1977)
		252--280}.
	
	\bibitem{Kovacs:1978eu}
	S.~J. Kovacs and K.~S. Thorne, ``{The Generation of Gravitational Waves. 4.
		Bremsstrahlung},''
	\href{http://dx.doi.org/10.1086/156350}{{\em Astrophys. J.} {\bf 224} (1978)
		62--85}.
	
	\bibitem{Jakobsen:2021smu}
	G.~U. Jakobsen, G.~Mogull, J.~Plefka, and J.~Steinhoff, ``{Classical
		Gravitational Bremsstrahlung from a Worldline Quantum Field Theory},''
	\href{http://dx.doi.org/10.1103/PhysRevLett.126.201103}{{\em Phys. Rev.
			Lett.} {\bf 126} (2021) no.~20, 201103},
	\href{http://arxiv.org/abs/2101.12688}{{\tt arXiv:2101.12688 [gr-qc]}}.
	
	\bibitem{Mougiakakos:2021ckm}
	S.~Mougiakakos, M.~M. Riva, and F.~Vernizzi, ``{Gravitational Bremsstrahlung in
		the post-Minkowskian effective field theory},''
	\href{http://dx.doi.org/10.1103/PhysRevD.104.024041}{{\em Phys. Rev. D} {\bf
			104} (2021) no.~2, 024041}, \href{http://arxiv.org/abs/2102.08339}{{\tt
			arXiv:2102.08339 [gr-qc]}}.
	
	\bibitem{DeAngelis:2023lvf}
	S.~De~Angelis, R.~Gonzo, and P.~P. Novichkov, ``{Spinning waveforms from KMOC
		at leading order},'' \href{http://arxiv.org/abs/2309.17429}{{\tt
			arXiv:2309.17429 [hep-th]}}.
	
	\bibitem{Brandhuber:2023hhl}
	A.~Brandhuber, G.~R. Brown, G.~Chen, J.~Gowdy, and G.~Travaglini, ``{Resummed
		spinning waveforms from five-point amplitudes},''
	\href{http://dx.doi.org/10.1007/JHEP02(2024)026}{{\em JHEP} {\bf 02} (2024)
		026}, \href{http://arxiv.org/abs/2310.04405}{{\tt arXiv:2310.04405
			[hep-th]}}.
	
	\bibitem{Aoude:2023dui}
	R.~Aoude, K.~Haddad, C.~Heissenberg, and A.~Helset, ``{Leading-order
		gravitational radiation to all spin orders},''
	\href{http://dx.doi.org/10.1103/PhysRevD.109.036007}{{\em Phys. Rev. D} {\bf
			109} (2024) no.~3, 036007}, \href{http://arxiv.org/abs/2310.05832}{{\tt
			arXiv:2310.05832 [hep-th]}}.
	
	\bibitem{Cristofoli:2021vyo}
	A.~Cristofoli, R.~Gonzo, D.~A. Kosower, and D.~O'Connell, ``{Waveforms from
		amplitudes},'' \href{http://dx.doi.org/10.1103/PhysRevD.106.056007}{{\em
			Phys. Rev. D} {\bf 106} (2022) no.~5, 056007},
	\href{http://arxiv.org/abs/2107.10193}{{\tt arXiv:2107.10193 [hep-th]}}.
	
	\bibitem{Goldberger:2016iau}
	W.~D. Goldberger and A.~K. Ridgway, ``{Radiation and the classical double copy
		for color charges},''
	\href{http://dx.doi.org/10.1103/PhysRevD.95.125010}{{\em Phys. Rev. D} {\bf
			95} (2017) no.~12, 125010}, \href{http://arxiv.org/abs/1611.03493}{{\tt
			arXiv:1611.03493 [hep-th]}}.
	
	\bibitem{Luna:2017dtq}
	A.~Luna, I.~Nicholson, D.~O'Connell, and C.~D. White, ``{Inelastic Black Hole
		Scattering from Charged Scalar Amplitudes},''
	\href{http://dx.doi.org/10.1007/JHEP03(2018)044}{{\em JHEP} {\bf 03} (2018)
		044}, \href{http://arxiv.org/abs/1711.03901}{{\tt arXiv:1711.03901
			[hep-th]}}.
	
	\bibitem{Brandhuber:2023hhy}
	A.~Brandhuber, G.~R. Brown, G.~Chen, S.~De~Angelis, J.~Gowdy, and
	G.~Travaglini, ``{One-loop gravitational bremsstrahlung and waveforms from a
		heavy-mass effective field theory},''
	\href{http://dx.doi.org/10.1007/JHEP06(2023)048}{{\em JHEP} {\bf 06} (2023)
		048}, \href{http://arxiv.org/abs/2303.06111}{{\tt arXiv:2303.06111
			[hep-th]}}.
	
	\bibitem{Herderschee:2023fxh}
	A.~Herderschee, R.~Roiban, and F.~Teng, ``{The sub-leading scattering waveform
		from amplitudes},'' \href{http://dx.doi.org/10.1007/JHEP06(2023)004}{{\em
			JHEP} {\bf 06} (2023)  004}, \href{http://arxiv.org/abs/2303.06112}{{\tt
			arXiv:2303.06112 [hep-th]}}.
	
	\bibitem{Elkhidir:2023dco}
	A.~Elkhidir, D.~O'Connell, M.~Sergola, and I.~A. Vazquez-Holm, ``{Radiation and
		Reaction at One Loop},'' \href{http://arxiv.org/abs/2303.06211}{{\tt
			arXiv:2303.06211 [hep-th]}}.
	
	\bibitem{Georgoudis:2023lgf}
	A.~Georgoudis, C.~Heissenberg, and I.~Vazquez-Holm, ``{Inelastic exponentiation
		and classical gravitational scattering at one loop},''
	\href{http://dx.doi.org/10.1007/JHEP06(2023)126}{{\em JHEP} {\bf 06} (2023)
		126}, \href{http://arxiv.org/abs/2303.07006}{{\tt arXiv:2303.07006
			[hep-th]}}.
	
	\bibitem{Cristofoli:2021jas}
	A.~Cristofoli, R.~Gonzo, N.~Moynihan, D.~O'Connell, A.~Ross, M.~Sergola, and
	C.~D. White, ``{The Uncertainty Principle and Classical Amplitudes},''
	\href{http://arxiv.org/abs/2112.07556}{{\tt arXiv:2112.07556 [hep-th]}}.
	
	\bibitem{DiVecchia:2022piu}
	P.~Di~Vecchia, C.~Heissenberg, R.~Russo, and G.~Veneziano, ``{Classical
		gravitational observables from the Eikonal operator},''
	\href{http://dx.doi.org/10.1016/j.physletb.2023.138049}{{\em Phys. Lett. B}
		{\bf 843} (2023)  138049}, \href{http://arxiv.org/abs/2210.12118}{{\tt
			arXiv:2210.12118 [hep-th]}}.
	
	\bibitem{Caron-Huot:2023vxl}
	S.~Caron-Huot, M.~Giroux, H.~S. Hannesdottir, and S.~Mizera, ``{What can be
		measured asymptotically?},''
	\href{http://dx.doi.org/10.1007/JHEP01(2024)139}{{\em JHEP} {\bf 01} (2024)
		139}, \href{http://arxiv.org/abs/2308.02125}{{\tt arXiv:2308.02125
			[hep-th]}}.
	
	\bibitem{Kosower:2018adc}
	D.~A. Kosower, B.~Maybee, and D.~O'Connell, ``{Amplitudes, Observables, and
		Classical Scattering},''
	\href{http://dx.doi.org/10.1007/JHEP02(2019)137}{{\em JHEP} {\bf 02} (2019)
		137},
	\href{http://arxiv.org/abs/1811.10950}{{\tt arXiv:1811.10950 [hep-th]}}.
	
	\bibitem{Bini:2023fiz}
	D.~Bini, T.~Damour, and A.~Geralico, ``{Comparing one-loop gravitational
		bremsstrahlung amplitudes to the multipolar-post-Minkowskian waveform},''
	\href{http://dx.doi.org/10.1103/PhysRevD.108.124052}{{\em Phys. Rev. D} {\bf
			108} (2023) no.~12, 124052}, \href{http://arxiv.org/abs/2309.14925}{{\tt
			arXiv:2309.14925 [gr-qc]}}.
	
	\bibitem{Georgoudis:2023ozp}
	A.~Georgoudis, C.~Heissenberg, and I.~Vazquez-Holm, ``{Addendum to: Inelastic
		exponentiation and classical gravitational scattering at one loop},''
	\href{http://dx.doi.org/10.1007/JHEP02(2024)161}{{\em JHEP} {\bf 2024} (2024)
		no.~02, 161}, \href{http://arxiv.org/abs/2312.14710}{{\tt arXiv:2312.14710
			[hep-th]}}.
	
	\bibitem{Bohnenblust:2023qmy}
	L.~Bohnenblust, H.~Ita, M.~Kraus, and J.~Schlenk, ``{Gravitational
		Bremsstrahlung in Black-Hole Scattering at $\mathcal{O}(G^3)$: Linear-in-Spin
		Effects},'' \href{http://arxiv.org/abs/2312.14859}{{\tt arXiv:2312.14859
			[hep-th]}}.
	
	\bibitem{Blanchet:1987wq}
	L.~Blanchet and T.~Damour, ``{Tail Transported Temporal Correlations in the
		Dynamics of a Gravitating System},''
	\href{http://dx.doi.org/10.1103/PhysRevD.37.1410}{{\em Phys. Rev. D} {\bf 37}
		(1988)  1410}.
	
	\bibitem{Blanchet:1992br}
	L.~Blanchet and T.~Damour, ``{Hereditary effects in gravitational radiation},''
	\href{http://dx.doi.org/10.1103/PhysRevD.46.4304}{{\em Phys. Rev. D} {\bf 46}
		(1992)  4304--4319}.
	
	\bibitem{Blanchet:1993ng}
	L.~Blanchet, ``{Time asymmetric structure of gravitational radiation},''
	\href{http://dx.doi.org/10.1103/PhysRevD.47.4392}{{\em Phys. Rev. D} {\bf 47}
		(1993)  4392--4420}.
	
	\bibitem{Weinberg:1965nx}
	S.~Weinberg, ``{Infrared photons and gravitons},''
	\href{http://dx.doi.org/10.1103/PhysRev.140.B516}{{\em Phys. Rev.} {\bf 140}
		(1965)  B516--B524}.
	
	\bibitem{Goldberger:2009qd}
	W.~D. Goldberger and A.~Ross, ``{Gravitational radiative corrections from
		effective field theory},''
	\href{http://dx.doi.org/10.1103/PhysRevD.81.124015}{{\em Phys. Rev. D} {\bf
			81} (2010)  124015}, \href{http://arxiv.org/abs/0912.4254}{{\tt
			arXiv:0912.4254 [gr-qc]}}.
	
	\bibitem{Porto:2012as}
	R.~A. Porto, A.~Ross, and I.~Z. Rothstein, ``{Spin induced multipole moments
		for the gravitational wave amplitude from binary inspirals to 2.5
		Post-Newtonian order},''
	\href{http://dx.doi.org/10.1088/1475-7516/2012/09/028}{{\em JCAP} {\bf 09}
		(2012)  028}, \href{http://arxiv.org/abs/1203.2962}{{\tt arXiv:1203.2962
			[gr-qc]}}.
	
	\bibitem{Blanchet:1985sp}
	L.~Blanchet and T.~Damour, ``{Radiative gravitational fields in general
		relativity I. general structure of the field outside the source},''
	\href{http://dx.doi.org/10.1098/rsta.1986.0125}{{\em Phil. Trans. Roy. Soc.
			Lond. A} {\bf 320} (1986)  379--430}.
	
	\bibitem{Blanchet:1986dk}
	L.~Blanchet, ``{Radiative gravitational fields in general relativity. 2.
		Asymptotic behaviour at future null infinity},''
	\href{http://dx.doi.org/10.1098/rspa.1987.0022}{{\em Proc. Roy. Soc. Lond. A}
		{\bf 409} (1987)  383--399}.
	
	\bibitem{Blanchet:1998in}
	L.~Blanchet, ``{On the multipole expansion of the gravitational field},''
	\href{http://dx.doi.org/10.1088/0264-9381/15/7/013}{{\em Class. Quant. Grav.}
		{\bf 15} (1998)  1971--1999}, \href{http://arxiv.org/abs/gr-qc/9801101}{{\tt
			arXiv:gr-qc/9801101}}.
	
	\bibitem{Poujade:2001ie}
	O.~Poujade and L.~Blanchet, ``{PostNewtonian approximation for isolated systems
		calculated by matched asymptotic expansions},''
	\href{http://dx.doi.org/10.1103/PhysRevD.65.124020}{{\em Phys. Rev. D} {\bf
			65} (2002)  124020}, \href{http://arxiv.org/abs/gr-qc/0112057}{{\tt
			arXiv:gr-qc/0112057}}.
	
	\bibitem{Blanchet:2013haa}
	L.~Blanchet, ``{Gravitational Radiation from Post-Newtonian Sources and
		Inspiralling Compact Binaries},''
	\href{http://dx.doi.org/10.12942/lrr-2014-2}{{\em Living Rev. Rel.} {\bf 17}
		(2014)  2}, \href{http://arxiv.org/abs/1310.1528}{{\tt arXiv:1310.1528
			[gr-qc]}}.
	
	\bibitem{Sahoo:2018lxl}
	B.~Sahoo and A.~Sen, ``{Classical and Quantum Results on Logarithmic Terms in
		the Soft Theorem in Four Dimensions},''
	\href{http://dx.doi.org/10.1007/JHEP02(2019)086}{{\em JHEP} {\bf 02} (2019)
		086},
	\href{http://arxiv.org/abs/1808.03288}{{\tt arXiv:1808.03288 [hep-th]}}.
	
	\bibitem{Sahoo:2021ctw}
	B.~Sahoo and A.~Sen, ``{Classical soft graviton theorem rewritten},''
	\href{http://dx.doi.org/10.1007/JHEP01(2022)077}{{\em JHEP} {\bf 01} (2022)
		077}, \href{http://arxiv.org/abs/2105.08739}{{\tt arXiv:2105.08739
			[hep-th]}}.
	
	\bibitem{Veneziano:2022zwh}
	G.~Veneziano and G.~A. Vilkovisky, ``{Angular momentum loss in gravitational
		scattering, radiation reaction, and the Bondi gauge ambiguity},''
	\href{http://dx.doi.org/10.1016/j.physletb.2022.137419}{{\em Phys. Lett. B}
		{\bf 834} (2022)  137419}, \href{http://arxiv.org/abs/2201.11607}{{\tt
			arXiv:2201.11607 [gr-qc]}}.
	
	\bibitem{DiVecchia:2023frv}
	P.~Di~Vecchia, C.~Heissenberg, R.~Russo, and G.~Veneziano, ``{The gravitational
		eikonal: from particle, string and brane collisions to black-hole
		encounters},'' \href{http://arxiv.org/abs/2306.16488}{{\tt arXiv:2306.16488
			[hep-th]}}.
	
	\bibitem{Bernard:2017bvn}
	L.~Bernard, L.~Blanchet, A.~Boh\'e, G.~Faye, and S.~Marsat, ``{Dimensional
		regularization of the IR divergences in the Fokker action of point-particle
		binaries at the fourth post-Newtonian order},''
	\href{http://dx.doi.org/10.1103/PhysRevD.96.104043}{{\em Phys. Rev. D} {\bf
			96} (2017) no.~10, 104043}, \href{http://arxiv.org/abs/1706.08480}{{\tt
			arXiv:1706.08480 [gr-qc]}}.
	
	\bibitem{Blanchet:2023bwj}
	L.~Blanchet, G.~Faye, Q.~Henry, F.~Larrouturou, and D.~Trestini,
	``{Gravitational-Wave Phasing of Quasicircular Compact Binary Systems to the
		Fourth-and-a-Half Post-Newtonian Order},''
	\href{http://dx.doi.org/10.1103/PhysRevLett.131.121402}{{\em Phys. Rev.
			Lett.} {\bf 131} (2023) no.~12, 121402},
	\href{http://arxiv.org/abs/2304.11185}{{\tt arXiv:2304.11185 [gr-qc]}}.
	
	\bibitem{Blanchet:2023sbv}
	L.~Blanchet, G.~Faye, Q.~Henry, F.~Larrouturou, and D.~Trestini,
	``{Gravitational-wave flux and quadrupole modes from quasicircular
		nonspinning compact binaries to the fourth post-Newtonian order},''
	\href{http://dx.doi.org/10.1103/PhysRevD.108.064041}{{\em Phys. Rev. D} {\bf
			108} (2023) no.~6, 064041}, \href{http://arxiv.org/abs/2304.11186}{{\tt
			arXiv:2304.11186 [gr-qc]}}.
	
	\bibitem{Bini:2021gat}
	D.~Bini, T.~Damour, and A.~Geralico, ``{Radiative contributions to
		gravitational scattering},''
	\href{http://dx.doi.org/10.1103/PhysRevD.104.084031}{{\em Phys. Rev. D} {\bf
			104} (2021) no.~8, 084031}, \href{http://arxiv.org/abs/2107.08896}{{\tt
			arXiv:2107.08896 [gr-qc]}}.
	
	\bibitem{Bini:2022enm}
	D.~Bini, T.~Damour, and A.~Geralico, ``{Radiated momentum and radiation
		reaction in gravitational two-body scattering including time-asymmetric
		effects},'' \href{http://dx.doi.org/10.1103/PhysRevD.107.024012}{{\em Phys.
			Rev. D} {\bf 107} (2023) no.~2, 024012},
	\href{http://arxiv.org/abs/2210.07165}{{\tt arXiv:2210.07165 [gr-qc]}}.
	
	\bibitem{Bini:2021jmj}
	D.~Bini and A.~Geralico, ``{Frequency domain analysis of the gravitational wave
		energy loss in hyperbolic encounters},''
	\href{http://dx.doi.org/10.1103/PhysRevD.104.104019}{{\em Phys. Rev. D} {\bf
			104} (2021) no.~10, 104019}, \href{http://arxiv.org/abs/2108.02472}{{\tt
			arXiv:2108.02472 [gr-qc]}}.
	
	\bibitem{Cho:2018upo}
	G.~Cho, A.~Gopakumar, M.~Haney, and H.~M. Lee, ``{Gravitational waves from
		compact binaries in post-Newtonian accurate hyperbolic orbits},''
	\href{http://dx.doi.org/10.1103/PhysRevD.98.024039}{{\em Phys. Rev. D} {\bf
			98} (2018) no.~2, 024039}, \href{http://arxiv.org/abs/1807.02380}{{\tt
			arXiv:1807.02380 [gr-qc]}}.
	
	\bibitem{Mishra:2015bqa}
	C.~K. Mishra, K.~G. Arun, and B.~R. Iyer, ``{Third post-Newtonian gravitational
		waveforms for compact binary systems in general orbits: Instantaneous
		terms},'' \href{http://dx.doi.org/10.1103/PhysRevD.91.084040}{{\em Phys. Rev.
			D} {\bf 91} (2015) no.~8, 084040},
	\href{http://arxiv.org/abs/1501.07096}{{\tt arXiv:1501.07096 [gr-qc]}}.
	
	\bibitem{Blanchet:1996wx}
	L.~Blanchet, ``{Energy losses by gravitational radiation in inspiraling compact
		binaries to five halves postNewtonian order},''
	\href{http://dx.doi.org/10.1103/PhysRevD.71.129904}{{\em Phys. Rev. D} {\bf
			54} (1996)  1417--1438}, \href{http://arxiv.org/abs/gr-qc/9603048}{{\tt
			arXiv:gr-qc/9603048}}. [Erratum: Phys.Rev.D 71, 129904 (2005)].
	
	\bibitem{Blanchet:1997ji}
	L.~Blanchet, ``{Quadrupole-quadrupole gravitational waves},''
	\href{http://dx.doi.org/10.1088/0264-9381/15/1/008}{{\em Class. Quant. Grav.}
		{\bf 15} (1998)  89--111}, \href{http://arxiv.org/abs/gr-qc/9710037}{{\tt
			arXiv:gr-qc/9710037}}.
	
	\bibitem{Peraro:2016wsq}
	T.~Peraro, ``{Scattering amplitudes over finite fields and multivariate
		functional reconstruction},''
	\href{http://dx.doi.org/10.1007/JHEP12(2016)030}{{\em JHEP} {\bf 12} (2016)
		030}, \href{http://arxiv.org/abs/1608.01902}{{\tt arXiv:1608.01902
			[hep-ph]}}.
	
	\bibitem{Adamo:2024oxy}
	T.~Adamo, R.~Gonzo, and A.~Ilderton, ``{Gravitational Bound Waveforms from
		Amplitudes},'' \href{http://arxiv.org/abs/2402.00124}{{\tt arXiv:2402.00124
			[hep-th]}}.
	
\end{thebibliography}
\end{document}